\documentstyle[12pt,aasms4]{article}

\lefthead{I.M. Schmoldt et al.}
\righthead{Density and velocity fields and \bet from IRAS PSC$z$}

\newcommand{\hm}{$\,h^{-1}{\rm Mpc} \ $}
\newcommand{\bet}{$\beta$ }

\def\vect(#1){{\bf #1}}
\def\vec2(#1){\stackrel{\rightarrow}{#1}} 
\font\bmit=cmmib10
\def\bdelta{\hbox{\bmit\char"0E}}
\def\bzeta{\hbox{\bmit\char"10}}
\def\bmu{\hbox{\bmit\char"16}}
\def\bomega{\hbox{\bmit\char"21}}
\newcommand{\be}{\begin{equation}}
\newcommand{\ee}{\end{equation}}
\def\kms{\ifmmode\,{\rm km}\,{\rm s}^{-1}\else km$\,$s$^{-1}$\fi}

\begin{document}

\title{On density and velocity fields and \bet\\
from the IRAS PSC$z$ survey}

\author{Inga M. Schmoldt\altaffilmark{1},
Veikko Saar\altaffilmark{1,2},
Prasenjit Saha\altaffilmark{1},
E. Branchini\altaffilmark{3},
G.P. Efstathiou\altaffilmark{4},
C.S. Frenk\altaffilmark{5},
O. Keeble\altaffilmark{6},
S. Maddox\altaffilmark{4},
R. McMahon\altaffilmark{4},
S. Oliver\altaffilmark{6},
M. Rowan-Robinson\altaffilmark{6},
W. Saunders\altaffilmark{7},
W.J. Sutherland\altaffilmark{1},
H. Tadros\altaffilmark{1}, and
S.D.M. White\altaffilmark{8}}

\altaffiltext{1}{Department of Physics (Astrophysics), Oxford
University, Keble Road, Oxford OX1 3RH, UK}

\altaffiltext{2}{Department of Physics, The University of Tokyo, Tokyo
113, Japan}

\altaffiltext{3}{Kapteyn Sterrewacht, Rijksuniversiteit Groningen,
Postbus 800, 9700 AV Groningen, The Netherlands}

\altaffiltext{4}{Institute of Astronomy, University of Cambridge,
Madingley Road, Cambridge CB3 OHA, UK}

\altaffiltext{5}{Department of Physics, University of Durham, South
Road, Durham DH1 3LE, UK}

\altaffiltext{6}{Imperial College of Science, Technology, and
Medicine, Blackett Laboratory, Prince Consort Road, London SW1 2EZ,
UK}

\altaffiltext{7}{Institute for Astronomy, University of Edinburgh,
Blackford Hill, Edinburgh EH9 3JS, UK}

\altaffiltext{8}{Max Planck Institut f\"ur Astrophysik,
Karl-Schwarzschild-Stra\ss e 1, 85740 Garching, Germany}

\begin{abstract}
We present a version of the Fourier Bessel method first introduced by
\cite{fish94} and \cite{zar95} with two extensions: (a) we amend the
formalism to allow a generic galaxy weight which can be constant 
rather than the more
conventional overweighting of galaxies at high distances, and 
(b) we correct for the masked
zones by extrapolation of Fourier Bessel modes rather than by cloning
from the galaxy distribution in neighbouring regions. We test the
procedure extensively on $N$-body simulations and find that it gives
generally unbiased results but that the reconstructed velocities tend
to be overpredicted in high-density regions.
Applying the formalism to the PSZ$z$ redshift catalog, we find that
$\beta = 0.7 \pm 0.5$ from a comparison of the reconstructed Local
Group velocity to the CMB dipole. From an anisotropy test of the
velocity field, we find that $\beta = 1$ CDM models models normalized
to the current cluster abundance can be excluded with 90\%
confidence.
The density and velocity fields reconstructed agree with the fields
found by \cite{branch98} in most points. We find a back-infall into the
Great Attractor region (Hydra-Centaurus region) but tests suggest that
this may be an artifact. We identify all the major clusters in our density
field and confirm the existence of some previously identified possible
ones.
\end{abstract}


\section{Introduction}

Redshift surveys provide the only possibility for determining the
three-dimensional density field of luminous matter which is crucial
for studies of mass concentrations, the power spectrum, dynamical
analyses to probe the relationship between dark and luminous matter
and many other areas of observational cosmology. However, the
relation between redshift space and real space, although given
straightforwardly by Hubble's law in the limit of large distances, is
distorted on smaller scales by galaxy peculiar velocities (a full
analysis of this was first given by \cite{ka87}). Hence
correcting for the distortions becomes the most important physical
problem associated with redshift surveys.
Several different types of these surveys are currently available,
varying widely in their depth, sky coverage, and sampling
density. Since this paper will be mainly concerned with analysis of
the density and velocity fields, we use the recently completed PSC$z$
survey, which is ideal for our purposes because of its large number of
galaxy redshift and its near-complete sky-coverage. Its depth is small
enough to render effects of space curvature and galaxy evolution
negligible in our calculations.

There are several methods of correcting for distortions to recover the
real space density and velocity fields.  These all use either linear
theory or the Zel'dovich approximation, and are therefore ultimately
very limited in their ability to reconstruct the high-density
regions.\footnote{However, this may change when techniques based on
the fully nonlinear variational method of \cite{peeb89} are
developed to deal with large redshift surveys. The most recent work in
this area is by \cite{sharpe}.} They can roughly be separated into two
types of methods:

Iterative methods: Since peculiar velocities are caused by
gravitational acceleration, the velocity field can be recovered from
the density field. Iterative methods use the redshift space density
field to calculate a peculiar velocity field which can then be used to
correct the density field distortions. The procedure is repeated until
the velocity field converges -- see \cite{papII} and \cite{ka91} for
slightly different versions of this method.

Basis function methods: When transforming the measured redshift space
density field into a combination of angular and radial basis
functions, the distortion is concentrated in the radial part and its
correction becomes an algebraic matrix problem. There are several
different versions of this approach: \cite{nu94} transform the angular
part into basis functions but express the radial part in differential
equations which they then solve numerically. Their method uses the
Zel'dovich approximation.  \cite{fish94} (FLHLZ in the following) and
\cite{zar95} transform both angular and radial parts into basis
functions, using a combination of spherical harmonics and spherical
Bessel functions.

Other approaches include \cite{ks91} and \cite{sb}.

Work on nonlinear corrections for the evolution of the power spectrum
includes \cite{peacock94}, \cite{fn96}, and \cite{th96}; a critical
discussion of these may be found in \cite{hc98}.

In this paper, we extend and apply the Fourier Bessel method which was
first
introduced by FLHLZ and \cite{zar95}. Their treatment of dynamics
using this expansion is new, but the idea of using spherical harmonics
and even spherical Bessel functions goes
back to \cite{peeb73}. It has been applied to
redshift surveys by \cite{sl93}, \cite{shll92}, and \cite{fsl94}. We will
follow the original method closely but introduce two extensions:
\begin{itemize}
\item Conventionally, the weight given to each galaxy
increases as the number density of galaxies decreases (as it does with
radius in a flux-limited survey). We generalise the formalism to allow
constant or any other weight, since we feel that care has to be taken
with the conventional procedure:
The purpose of this type of weighting is to exaggerate the mass
of galaxies at higher radii where the sampling is poor, thereby making
it possible to determine a density field. Note however, that since this also
exaggerates the shot noise, the density field at high radii is
subsequently considered unreliable and fluctuations are smoothed
away.
\item We correct for the mask by also using a basis function approach
rather than the more usual cloning mechanism. The difference between
these two methods is not an issue for IRAS-based surveys, which have
very good sky coverage, but will be more interesting in the future
with the advent of very deep surveys of small angular coverage such as
the 2dF survey.
\end{itemize}

For our analysis, we use the recent PSC$z$ redshift survey.
It contains approximately 15,500 galaxies (almost all) detected in the IRAS
Point Source Catalog (\cite{ws96}, \cite{ws98})
with 60 $\mu$m flux
larger than 0.6 Jy. Our subsample contains 10549 PSC$z$ objects within
170 \hm and with positive galaxy identification. Regions not surveyed
by IRAS (two thin strips in ecliptic longitude and the area near the
galactic plane defined by a V-band extinction of $>$ 1.5 mag) are
excluded from the catalog which therefore covers $\sim$ 84 \% of the
sky.

The structure of the paper will be as follows: In section
\ref{sec_meth}, we describe the correction for redshift space
distortions using the Fourier-Bessel set of basis functions, and
discuss the Wiener Filter smoothing procedure for suppressing shot
noise.  Details are given in the Appendices. In all of this, we mostly
follow FLHLZ and \cite{zar95}, but we have chosen to give derivations
in full since we extended the original formalism and also changed some
normalisations to render them more intuitive.  In section
\ref{sec_maskcorr} we present our method of correcting the masked
areas. In section \ref{sec_sims} we then test the method on $N$-body
simulations to evaluate errors and systematic biases.

Sections \ref{sec_res1}, \ref{sec_res2}, and \ref{sec_res3}
present the analysis of the PSC$z$ redshift survey. In
section \ref{sec_res1}, we recover \bet from a comparison between the
reconstructed velocity of the Local Group and its value known from the
dipole anisotropy in the Cosmic Microwave Background (CMB). This
procedure has a long tradition since the Local Group velocity is one
of the few that is known accurately enough for a meaningful comparison
to be made to its reconstructed value. However, the
one-number-statistic nature of the procedure also makes it highly
susceptible to systematic biases and we therefore take
particular care to evaluate the error by analysing the scatter in the
$N$-body simulations.

Section \ref{sec_res2} analyses the magnitude of the redshift space
distortions to recover another estimate of \bet. It was first pointed
out by \cite{ka87} that the distortions themselves obviously depend on
\bet so that by analysing their magnitude, it should be possible to
recover a value for that parameter. Several versions of such analyses
have been done to date -- ours relies on the fact that, if we correct
the redshift space distortions assuming a wrong value for \bet, the
resulting density and velocity fields will be anisotropic, i.e., there
will be a systematic difference between the radial and the other two
directions. We develop a simple test for this anisotropy and estimate
\bet for the PSC$z$ again by comparing to the corresponding results
for $N$-body simulations.

In section \ref{sec_res3} we discuss the reconstructed density and
velocity fields respectively and compare them to other recent
reconstructions using the PSC$z$ and similar catalogs. Both of these
fields result naturally from the reconstruction method we use if in a
somewhat smoothed form due to the fact that our formalism only works
in the linear regime.

\section{The Fourier-Bessel method}
\label{sec_meth}
The idea of this method is to express the overdensity as a
Fourier-Bessel expansion
\be
\delta(r,\bomega) = \sum_{lmn} Y_{lm}(\bomega) j_l(k_{ln}r) \delta_{lmn}.
\label{eq_exp}
\ee
Here $\bomega$ denotes the angular coordinates, $Y_{lm}$ are spherical
harmonics, $j_l$ are spherical Bessel functions, $k_{ln}$ are a set of
wavenumbers that depend on the boundary conditions assumed, and
$\delta_{lmn}$ are the expansion coefficients.  Once the expansion
coefficients are known,\footnote{Since the expansion has to be finite,
the radial and angular resolutions are finite. The angular resolution
is given by the number of angular modes $l_{\rm max}$ and the radial
resolution by the number of radial modes $n_{\rm max}$. We want to
keep the resolution constant, so we have chosen to link $l$ and
$n_{\rm max}(l)$ such that $n_{\rm max}(l) + l/2 = R/r_{\rm min} $;
$r_{\rm min}$ is then the smallest scale probed by any given
mode. This is equivalent to the scheme used by FLHLZ; because the
zeroes of the Bessel function $j_l(z)$ are asymptotically given by
$z_{ln} \simeq \pi (n + l/2)$, their scheme of setting a fixed upper
limit to $z$ for every $j_l$ amounts to keeping $n_{\rm max}(l) + l/2$
a constant for every $l$.}.
the linear theory velocity field is easily calculated in terms of
these, as
\be \vect(v)(r,\bomega) = H_0\beta \sum_{lmn} \delta_{lmn}
\vect(\nabla) \left( Y_{lm}(\bomega) \frac{j_l(k_{ln}r)}{k_{ln}^2}
\right).
\ee

The problem is to determine the $\delta_{lmn}$ from redshift survey
data, especially correcting for distortions of redshift space
$(s,\bomega)$ relative to real space $(r,\bomega)$, arising from the
velocity field.  In this section we briefly describe how this is done,
in the method introduced by FLHLZ, which we extend and apply in this
paper. Full derivations are given in the Appendices.

\subsection{Inverting redshift space distortions}

From an all-sky redshift survey $(s_i,\bomega_i)$, one can compute
sums of the type
\be
\rho^S_{lmn} = \sum_{s_i<R} w(s_i) Y^*_{lm} (\bomega_i) j_l(k_{ln}s_i).
\label{eq_sdens}
\ee
Here $w(s)$ is a weighting function for the galaxies, which we allow
to be arbitrary. Now the $\rho^S_{lmn}$ seem like Fourier-Bessel
expansion coefficients for the density field in redshift space (hence
superscript $S$) but they depend on the selection function of the
survey---denoted by $\phi(r)$---through the sum, as well as on $w(s)$.
They are not the expansion coefficients unless $w(r)\phi(r)=1$.
In linear theory the $\rho^S_{lmn}$ can be related to the
sought-after $\delta_{lmn}$ by a messy but linear relation.
\be
{1\over\bar\rho} \sum_{n'} (P_l)^{-1}_{nn'} \rho_{lmn'}^{S,w}
- O_{lmn} = \sum_{n'} Z_{lnn'} \delta_{lmn'}^{Re}.
\label{eq_short}
\ee
Here
\be
O_{lmn} = \sqrt{4\pi} \int w \phi j_0(k_{ln} r) r^2 dr 
\label{eq_monop}
\ee
represents a monopole correction: initially we expand the density
field. In order to transform to the overdensity field, we have to
divide by the mean density $\bar{\rho}$ and subtract the $l=0$
(monopole) term from the coefficients. Note that the monopole
correction does not really depend on $m$, but we have kept the
index for consistency.  The second step in equation \ref{eq_short}
corrects the redshift space overdensity coefficients for the redshift
space distortions expressed by
\begin{eqnarray}
Z_{lnn'} &=& \int w \phi j_l(k_{ln} r) j_l(k_{n'} r) r^2 dr \nonumber \\
&&- \beta \int w \phi j_l(k_{ln}r) \left[ 
\left( \frac{l(l+1)}{k_{n'}^2 r^2} - 1 \right) j_l(k_{n'} r)
+ \frac{
j_l'(k_{n'} r)}{k_{n'} r} \frac{d\ln \phi}{d \ln r} \right] r^2 dr.
\end{eqnarray}
The matrices
\be
(P_l)^{-1}_{nn'} = \int w\phi j_l(k_{ln} r) j_l(k_{ln'}r) r^2 dr.
\label{eq_refw}
\ee
carry information about the weight function $w$.

Extracting $\delta_{lmn}$ from equation (\ref{eq_short}) involves
solving a matrix equation for each $(l,m)$. $P_{lnn'}$, $O_{lnn'}$,
and $Z_{lnn'}$ depend on the weight function $w(r)$ and the selection
function $\phi(r)$ but not on data.

The weight function $w(r)$ could be set to $1/\phi(r)$ to eliminate
the selection function from the formulas. The correction matrices
$P_{lnn'}$ would then become diagonal and (\ref{eq_refw}) would reduce
to the orthogonality relation (\ref{eq_ortho}) for spherical Bessel
functions.  The other extreme is to weight all galaxies equally $w(r)
= 1$, leaving the correction of the selection effect fully to the
matrices $P_{lmn}$. The difference between these two approaches lies
in the errors induced by shot noise.  The choice $w(r)=1$ tends
to extrapolate information from the well-sampled regions into
less-sampled regions, while also propagating shot noise from
less-sampled regions to well-sampled regions.  The choice
$w(r)=1/\phi(r)$ keeps the effect of shot noise more local. For the
data set we have analyzed in this paper, the difference made by $w(r)$
is very small (see figure \ref{fig_4} below) indicating that the
errors are nowhere dominated by shot noise.  However, the situation is
different when we encounter an analogous problem in Section
\ref{sec_maskcorr}.  There the selection function is angular and
becomes zero in the unobserved region; the information-propagation
aspect then becomes crucial for filling in this region.

The expressions above are equivalent to FLHLZ but not identical
because
\begin{itemize}
\item we have chosen to normalise the Fourier Bessel coefficients
so that $F_{lmn}$ always have the same dimensions as $F(r,\bomega)$.
\item our coefficients always refer to the Fourier Bessel basis set,
whereas in FLHLZ they sometimes refer to an intermediate basis set that
depends on the galaxy weight $w(r)$.
\item we work entirely in the CMB rather than in the Local Group rest
frame.
\end{itemize}
For this reason we have given full derivations in Appendix \ref{app_dyn}.

\subsection{Wiener Filter}
\label{sec_appl}
A Fourier-Bessel expansion computed as above will contain spurious
extra power from shot noise. This spurious power can be suppressed
by a Wiener Filter as
explained in \cite{numrec}, p.~548, where the filter is 
\be
\Phi  = \frac{\mbox{power in signal}}{\mbox{power in (signal +
noise)}}.
\label{eq_filtform}
\ee
The derivation of this expression, however, as given in \cite{numrec},
is only valid for a scalar transfer function, whereas in our case the
transfer function is $Z_{lnn'}$. We will therefore have to derive our
own Wiener Filter operator.  The filter and its derivation are given in
Appendix \ref{app_wf}, which mostly follows \cite{zar95}. But we have
presented the derivation in full since some of our normalizations are
different and since we needed to preserve the arbitrariness in $w(r)$
throughout.

\begin{figure}
\epsscale{0.5}
\plotone{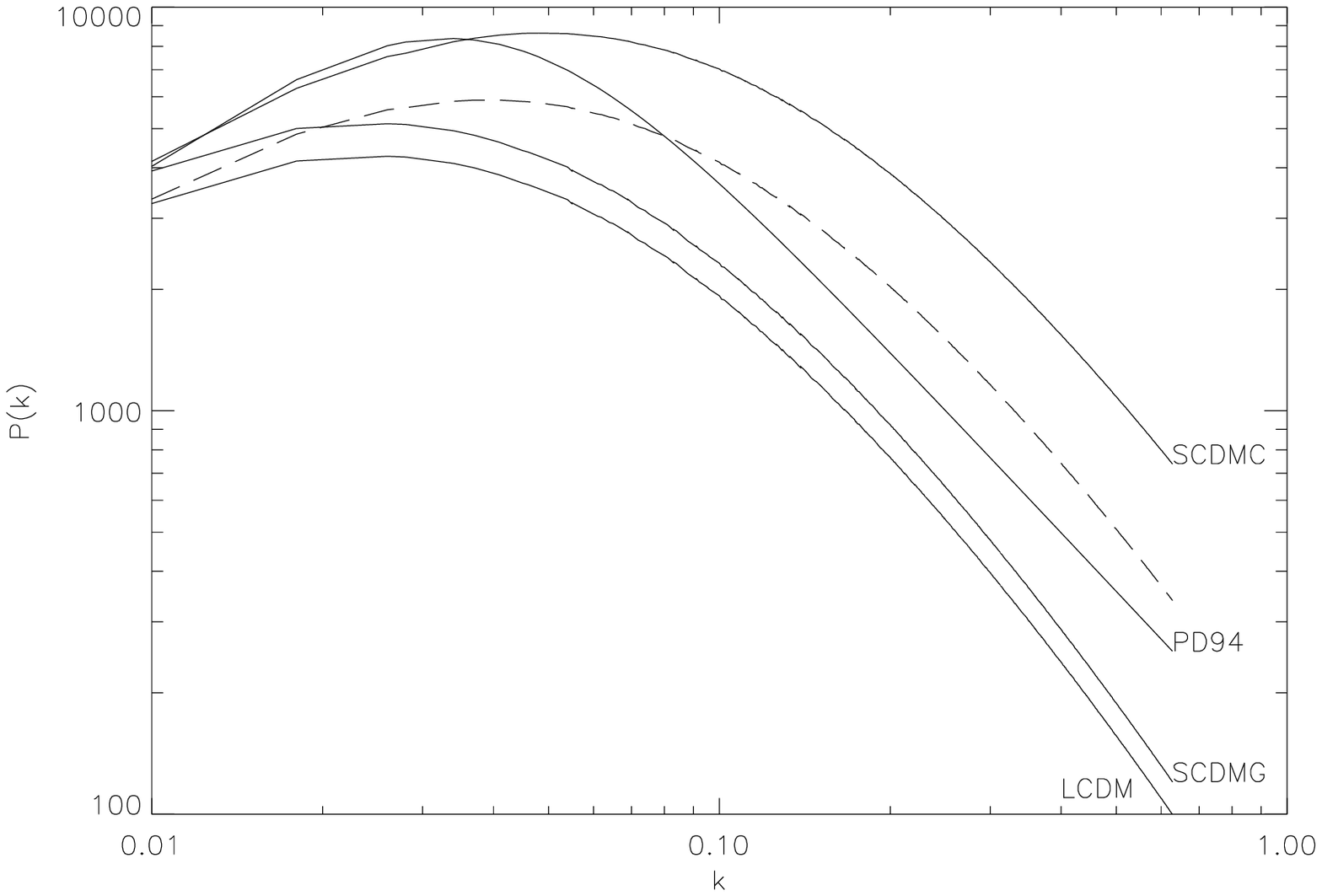}
\caption{Power spectra for all models considered: SCDMC (standard CDM with 
$\Gamma = 0.5$, $\sigma_8 = 1.1$, $\Omega = 1.0$), SCDMG (standard
CDM, COBE normalised with $\Gamma = 0.25$, $\sigma_8 = 0.55$,
$\Omega = 1.0$), LCDM ($\Lambda$ CDM  model with $\Gamma = 0.25$,
$\sigma_8 = 0.93$, $\Omega = 0.3$, $\Lambda = 0.7$), and PD94 (Peacock
\& Dodds, 1994), dashed line shows the power spectrum used for our
filtering ($\Gamma = 0.4$, $\sigma_8 = 0.8$)}
\label{fig_1}
\end{figure}

The Wiener Filter requires knowledge of the power spectrum $P(k)$ of
the underlying density fluctuations.  We do not know this in general,
but since the filter is a correction, a first order error in the
filter will only introduce a second order error in the full
reconstruction (cf \cite{numrec}, p.~548): even a fairly crude
approximation of the power spectrum will make the filter work. We
therefore use a CDM power spectrum with $\Gamma = 0.4$, normalised to
$\sigma_8 = 0.8$, as the filtering power spectrum for all our
reconstructions. This resembles the spectrum fitted to data by
\cite{peacock94} and all of the CDM spectra of the $N$-body
simulations (see figure \ref{fig_1}).  A Wiener Filter on the basis of
this power spectrum does as well in reconstructing velocities from
simulated catalogs as a filter based on the power spectrum
underlying the simulation in question.  The advantage of using this
constant filter rather than the `correct' filter based on the
underlying power spectrum is that the tests on the simulated
catalogs will then accurately reflect the error introduced into the
PSC$z$ reconstruction since we do not know the correct power spectrum
in that case.

Figure \ref{fig_1} shows power spectra for several cosmologies, including our
selected power spectrum and the one fitted to existing data by
\cite{peacock94}.

\section{Mask Correction}
\label{sec_maskcorr}
As mentioned above, the formalism developed in the preceding sections
is only valid for a full-sky catalog. Most redshift surveys are
masked in some way, so it is necessary to correct for the unobserved
parts of the sky. This is done by introducing fake galaxies into the
unobserved regions while trying to preserve the statistical
properties of the galaxy distribution of the observed
regions. The process is referred to as `mask correction'.

The usual method is to clone the fake from the observed galaxies by
extrapolating the distribution from the observed into the unobserved
regions.  This works reasonably well, but has the following
disadvantages:
\begin{itemize}
\item The joining-at-the-seams between the masked and unmasked regions
has a tendency to introduce spurious power on some scales (since the
correlation vanishes at the boundaries), which can
contaminate the Fourier Bessel coefficients.
\item The cloning method only works if the masked areas are very small
compared to the unmasked areas. It is hopeless for redshift surveys
with very small sky coverage such as the 2dF survey. It is therefore
interesting to try out a Fourier Bessel based mask correction here for
possible application with that survey. Naturally, with small sky
coverage, only high-$l$ modes might be usefully constrained, but this
may still extract information about density and velocity fields on
scales smaller than the survey.

\end{itemize}

As mentioned in section \ref{sec_dyn}, it is in principle
possible to treat the angular window function $\varphi(\bomega)$ at the
same time as the selection function $\phi(r)$ but it is
computationally problematic. We therefore choose to correct in two
steps: we first treat the mask and then the selection function. To
this end, we have to first expand the density field in such a way as
to decouple the angular modes completely from the radial modes.
Let $\varrho(s,\bomega)$ be the `raw' redshift space density, i.e.,
\be
\varrho(s,\bomega) = \phi(r) \rho^S(\vect(s)),
\ee
separate the angular and radial parts and expand
\be
\varrho(s,\bomega) = \sum_{lmn} \varrho_{lmn} Y_{lm}(\bomega)
j_0(k_{0n}s).
\ee
Note that because only $j_0$ occurs in this expression, 
the radial and angular basis functions are
truly independent and are not eigenfunctions of the Laplacian
operator. They do, however, form an adequate description of the
density field for the purpose of correcting for the angular mask. 

We minimise 
\be
\int \varphi(\bomega) \left| \varrho(s,\bomega) - \sum_{lmn} \varrho_{lmn}
Y_{lm}(\bomega) j_0(k_{0n}s) \right|^2 d\bomega s^2 ds
\ee
to obtain
\be
\int \varphi(\bomega) Y^*_{lm} j_0(k_{0n}s) d\bomega s^2 ds = C^{-1}_{0n}
\sum_{ll'mm'} \varrho_{l'm'n} 
\int \varphi(\bomega) Y_{l'm'}(\bomega) Y^*_{lm}(\bomega)
d\bomega.
\ee
The integrals can be replaced by sums over galaxies as before. We
invert the matrix on the right hand side and thereby recover
$\varrho_{lmn}$ and hence $\varrho(s,\bomega)$. This recovered density
field can then be sampled in the masked regions to provide the fake
galaxies.

As an additional way of minimising the errors in this procedure,
we invert the matrix with a conditioned inversion,
i.e., after diagonalising we note all those eigenvalues
which are less than 1 \% of the largest eigenvalue and eliminate the
corresponding eigensubspaces. The eigenvalues of the matrix
will be unity for unmasked modes and small for mostly masked
modes. Hence, the noise can be suppressed by suppressing the masked
modes. The procedure is similar to applying a Wiener Filter 
and ensures that the density field is
not dominated by spurious features in the masked zones.

\begin{figure}
\epsscale{1}
\plotone{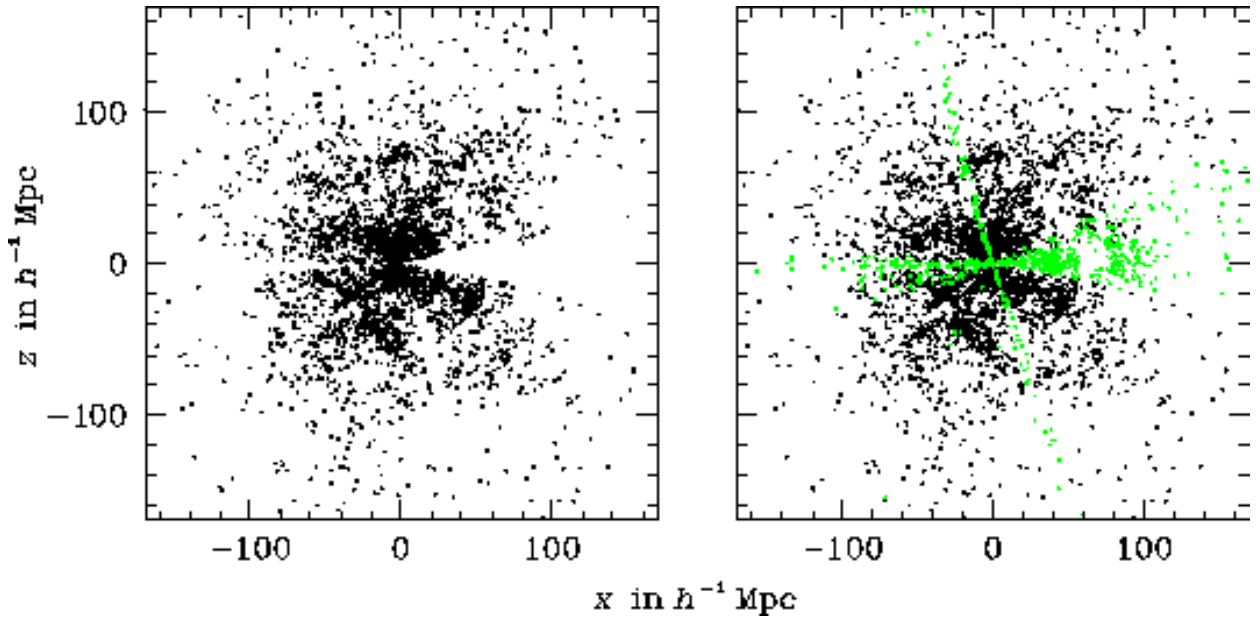}
\caption{ Projections of galaxy positions onto a plane in $x$-$z$ (where
the $x$-axis points towards $l=0^{\circ}$ and the $z$-axis towards
$b=90^{\circ}$; $y \mbox{ }\epsilon \mbox{ } [-20,20]$); unfilled and 
filled masked zones; Note how the mask correction produces voids
around (70,0) and (-20,50) which appears to be extrapolations of 
voids in the surrounding data.}
\label{fig_2}
\end{figure}

Figure \ref{fig_2} shows a slice of the
masked and unmasked galaxies in the 
$x$-$z$ plane.

\section{Tests on Simulated Catalogs}
\label{sec_sims}
The method described above can now be tested on simulated catalogs
with known selection function $\phi(r)$. We reconstruct the real space
density and velocity fields as described above and then sample it at
the real space galaxy positions. The $r_i$ of each galaxy is
reconstructed from its $s_i$ and the radial velocity field (cf
equation \ref{eq_vel1}) by
Newton-Raphson iteration. Reconstruction is done for a
radius $R$ of 170 \hm and a minimum radial resolution $r_{\rm min}$ of
5 \hm. Our galaxy weight is given by $w(r) = 1/\phi(r)$. We will use
this to be more in line with standard procedure in all of the following
reconstructions unless otherwise stated.

\begin{figure}
\epsscale{0.3}
\plotone{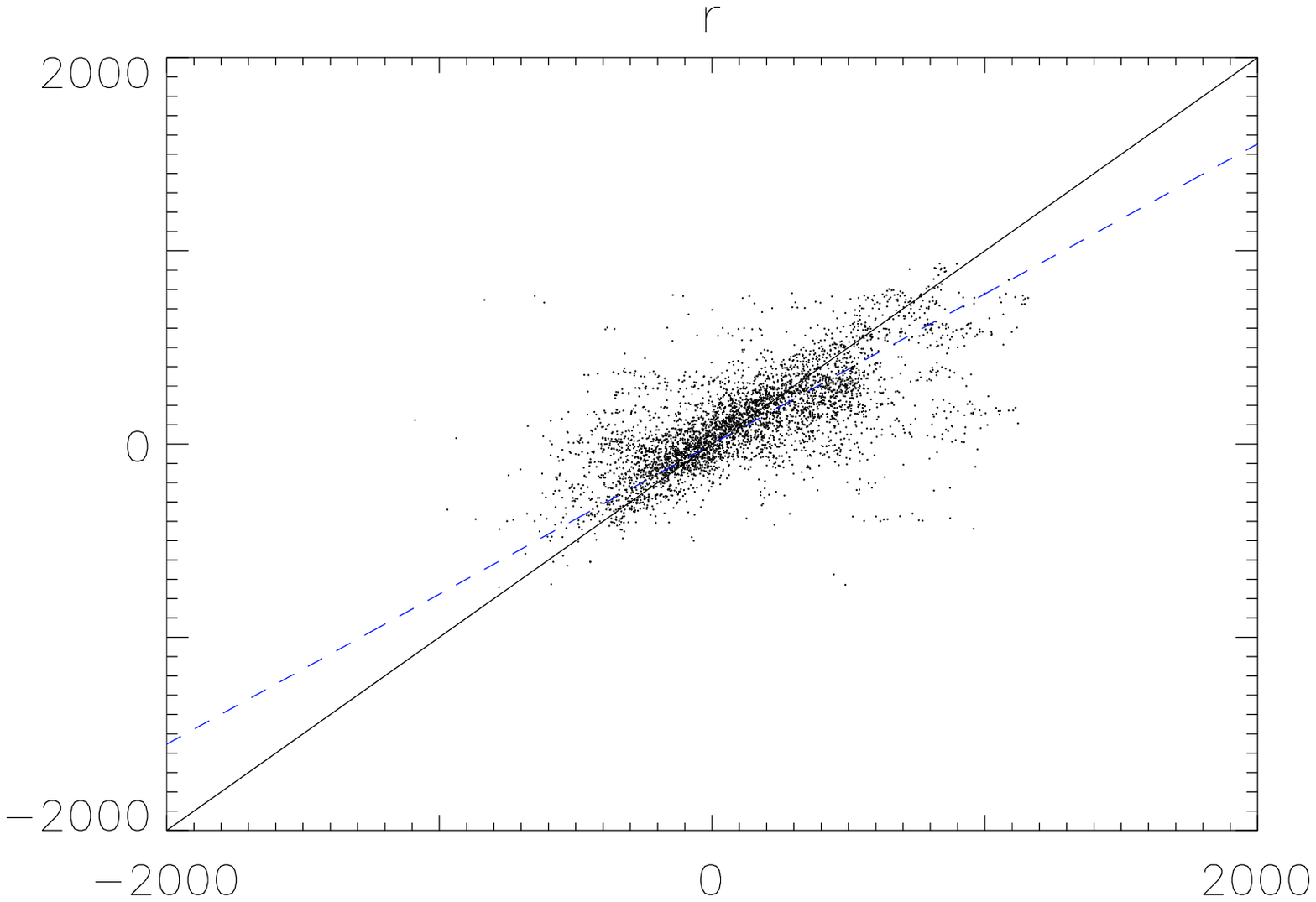}
\plotone{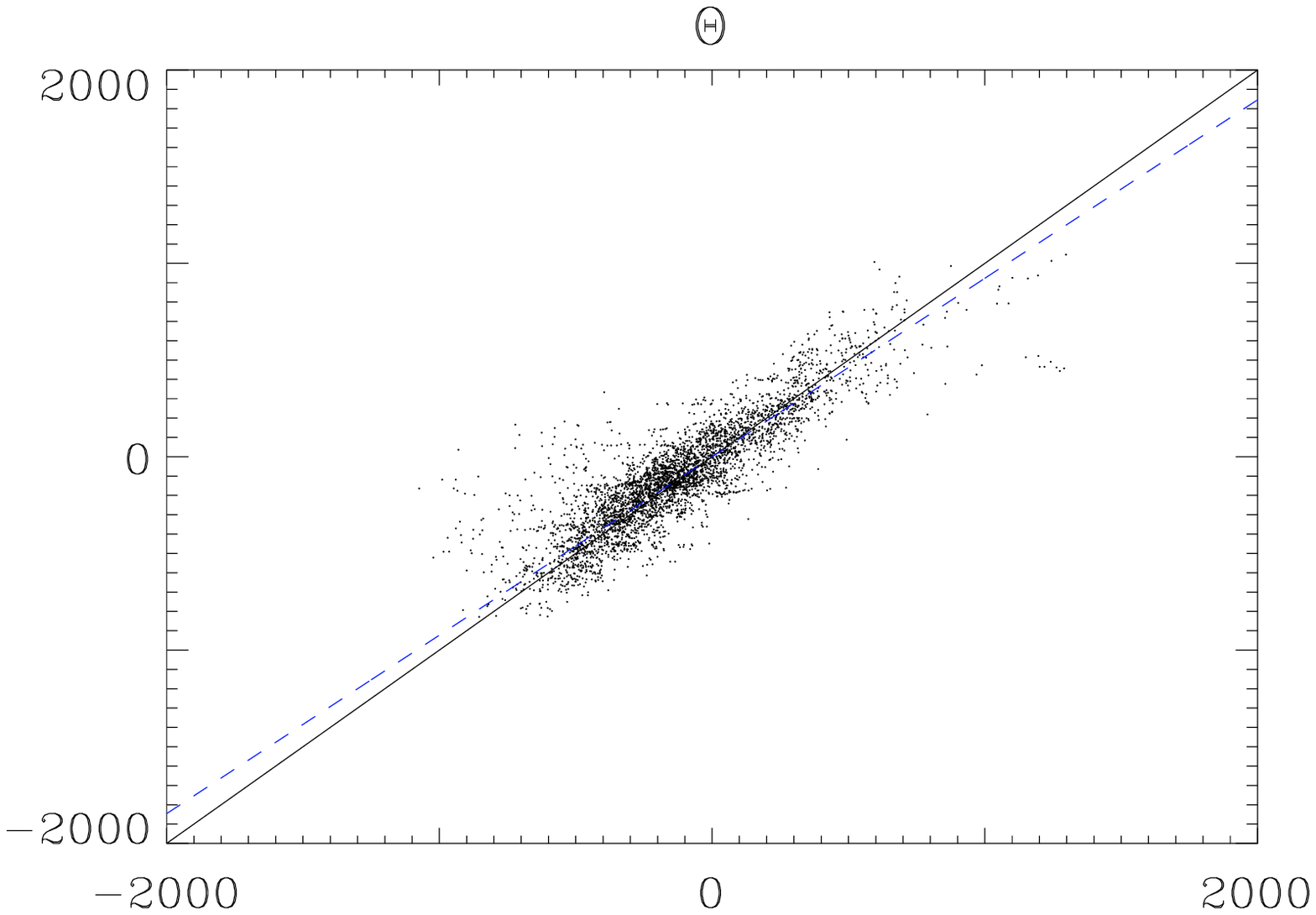}
\plotone{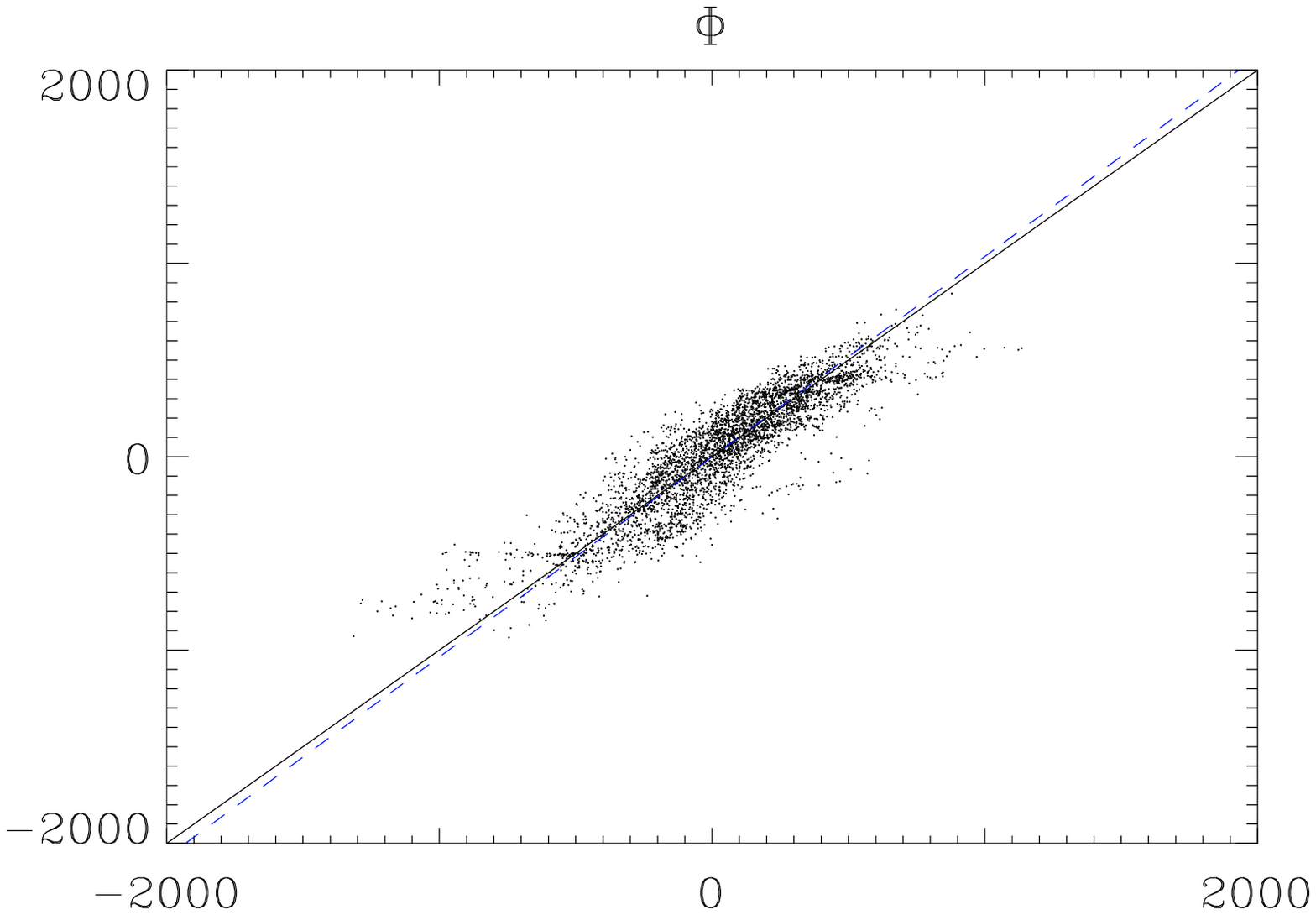}
\caption{Real (vertical) against reconstructed (horizontal) velocities: 
$v_r, v_{\theta}, v_{\phi}$ for
an LCDM catalog; plot galaxies out to 60 \hm, $1/\phi$ weighting,
constant filter; it is obvious that the
maximum error is in $v_r$.}
\label{fig_3}
\end{figure}

Figure \ref{fig_3} shows a comparison of real against predicted
velocities in the $r, \theta, \phi$ directions for an LCDM
catalog (for the parameters relating to the model abbreviations
LCDM, SCDMC, and SCDMG see the caption of \ref{fig_1}). 
The solid line is the line of perfect correlation and the
dashed line represents a fit to the data.  Note that this is not a fit
in the least squares sense. We found that least squares fits have the
tendency to be too easily dominated by outliers and therefore devised
a more stable fitting routine (see also the discussion in
\cite{numrec}, p. 700).  We constrain the fitted line to pass through
the origin and choose the slope to be such that if we were to rotate
the line by $90^{\circ}$ it would cross exactly half the points. This
does not ignore the outliers but it only accounts for their numbers
rather than for their distance to the line.

The figure illustrates that most of the error in the reconstruction
lies in the $r$ direction, as would be expected since the redshift
space distortions reside exclusively in that dimension.  The error in
the $\theta$ and $\phi$ directions is produced by shot noise only.  In
the following we will here only plot the radial velocity comparisons,
since they are the most interesting in assessing the performance of
the method.  Note also that the reconstruction was performed using a
`constant filter', i.e., instead of using the correct power spectrum
for this cosmology, we used the chosen power spectrum discussed in
section \ref{sec_appl}.

\begin{figure}
\epsscale{0.3}
\plotone{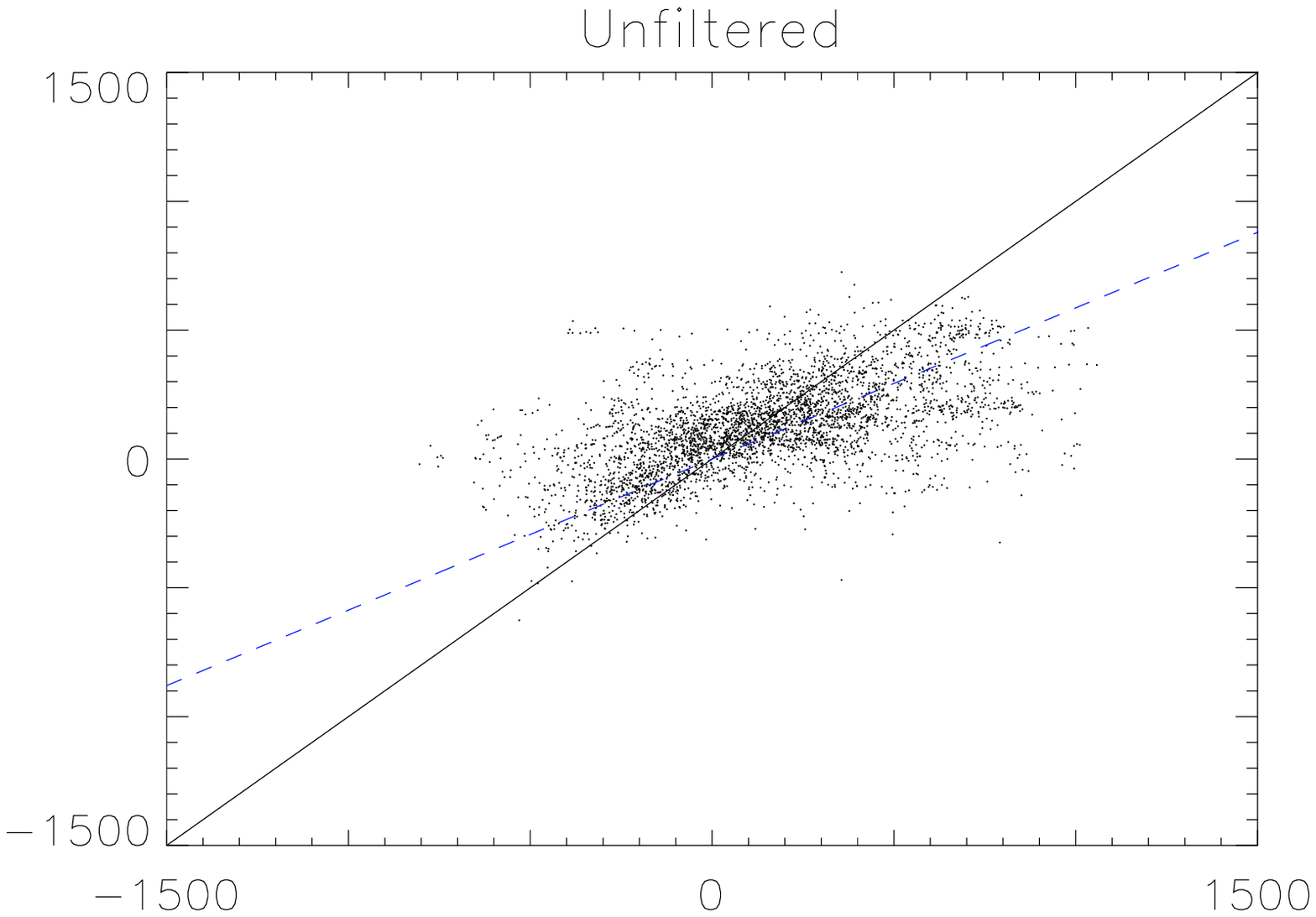}
\plotone{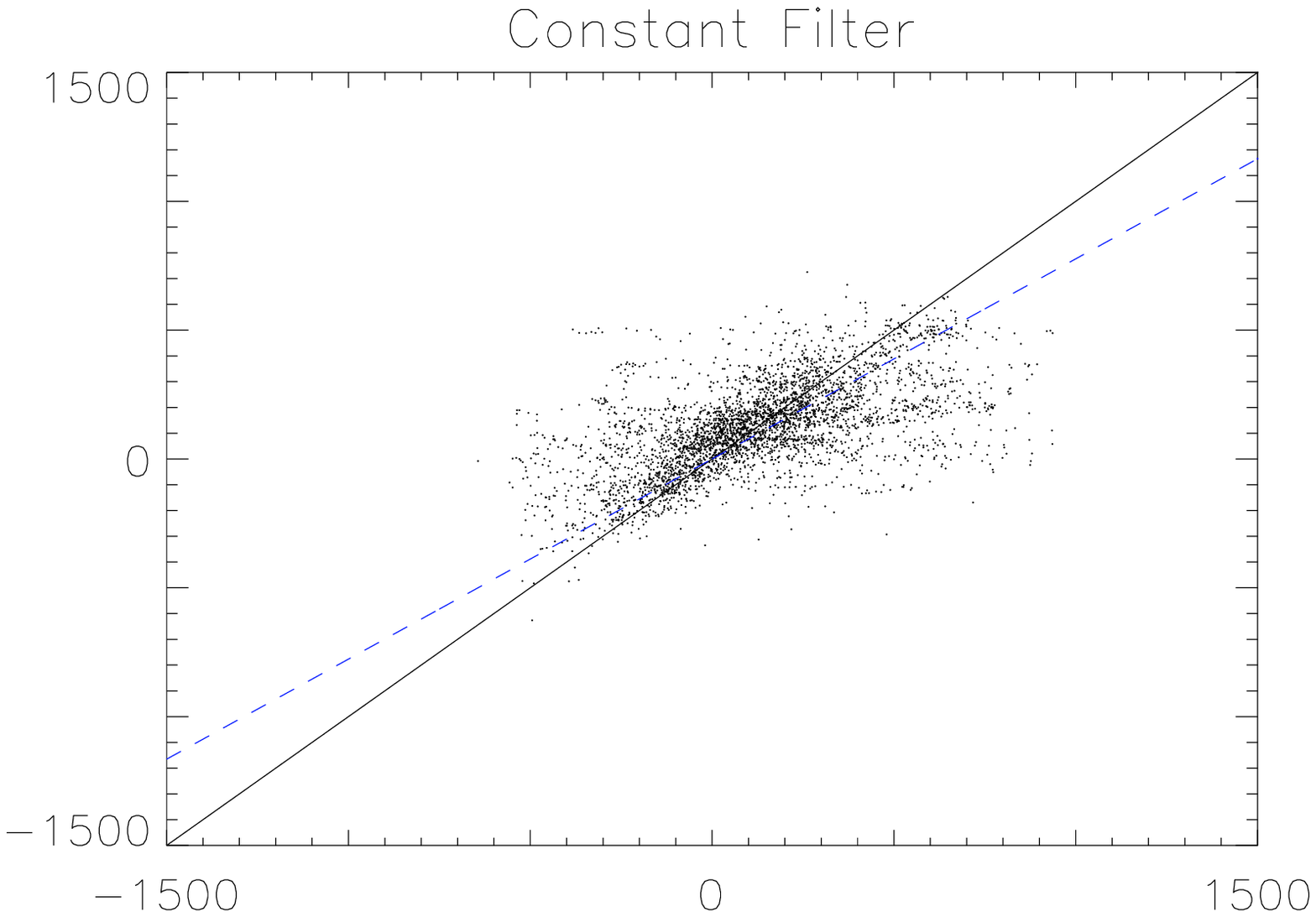}
\plotone{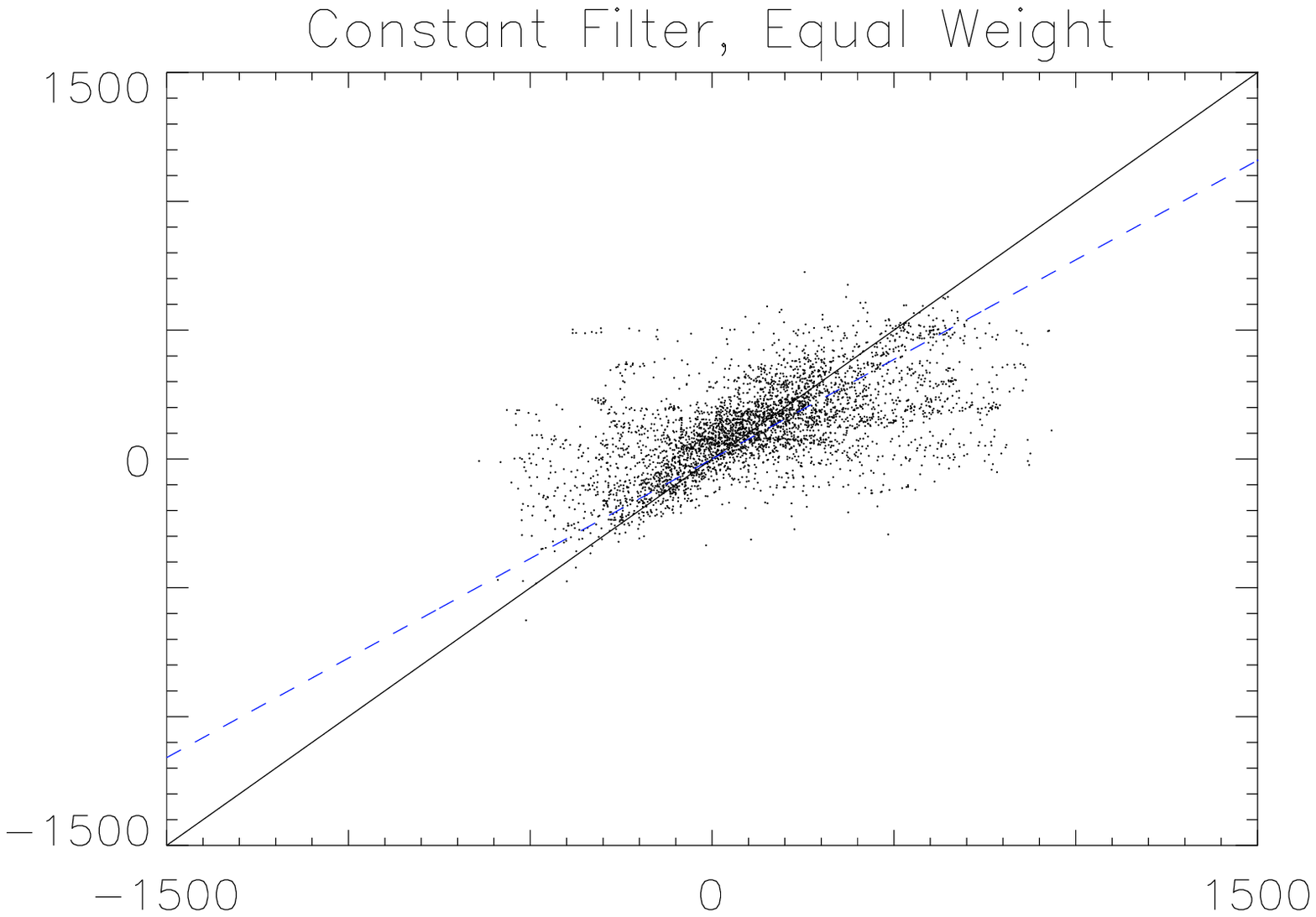}
\caption{ Real against predicted velocities for three different types of
filters; LCDM catalog; plot galaxies out to 60 \hm, $1/\phi$
weighting in all but the last panel.}
\label{fig_4}
\end{figure}

Figure \ref{fig_4} shows the effect of using different types of filter. The
first panel shows a reconstruction for an LCDM catalog using no filter.
The second panel shows the same catalog, but in this case the velocities
were reconstructed with the constant filter. The variance is smaller
now since most of the shot noise is smoothed away.
The third panel shows exactly the same case but we have here used a
constant weighting function $w(r) = 1$. The result illustrates 
that in the case of the velocity reconstruction the results
are the same whether we use an inverse selection function or a
constant weighting for each galaxy.

{\vskip 10pt}

\begin{figure}
\epsscale{0.3}
\plotone{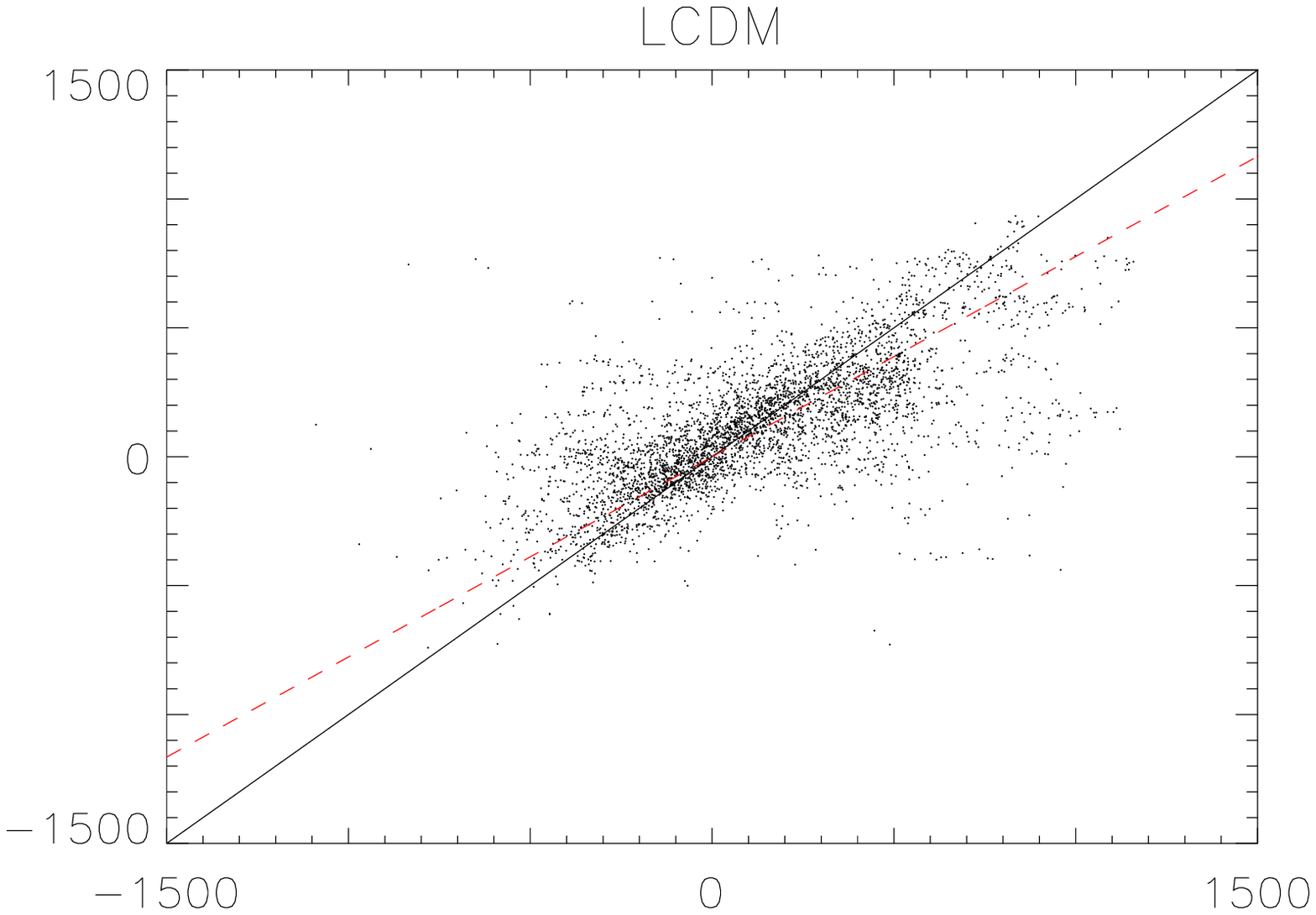}
\plotone{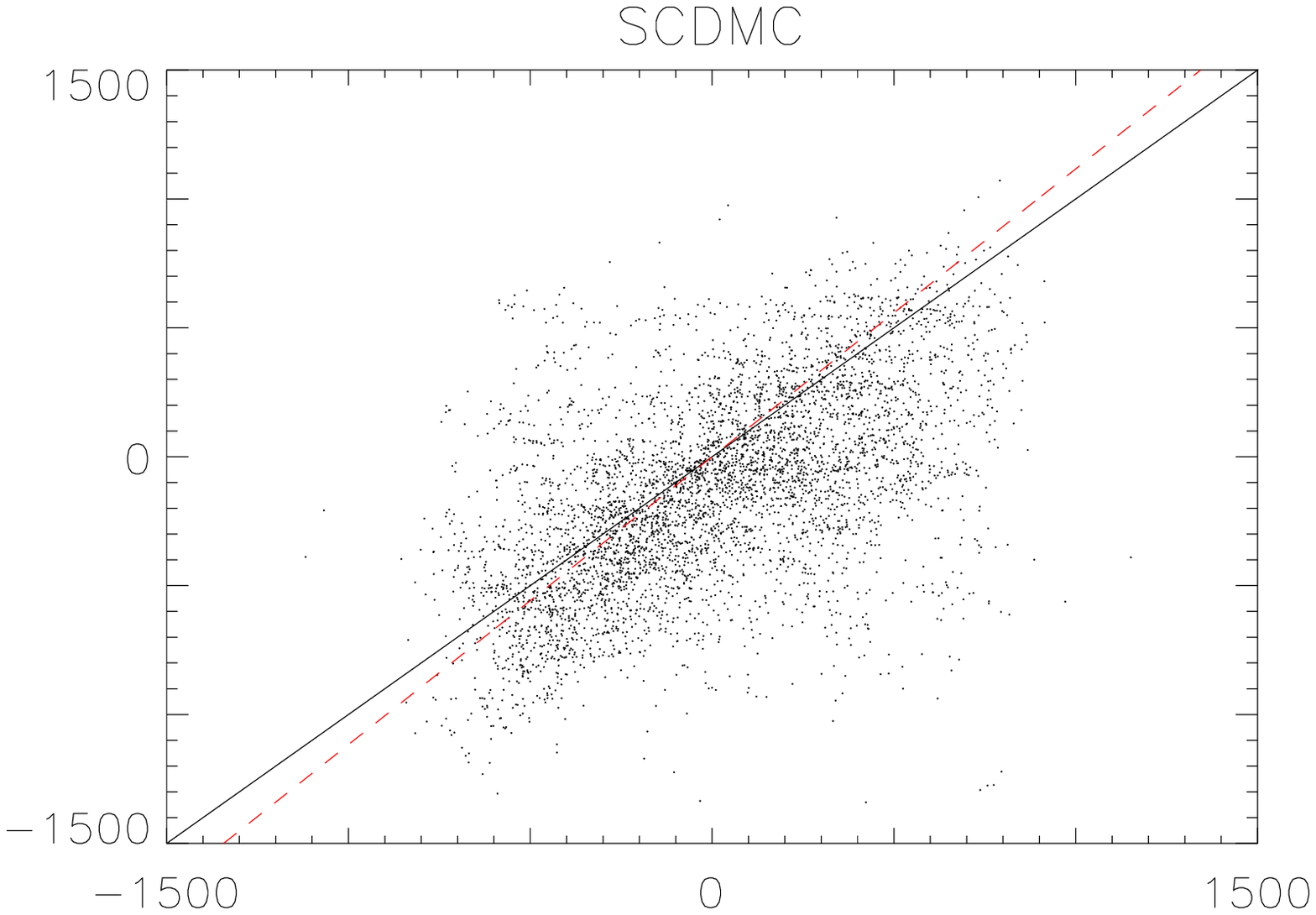}
\plotone{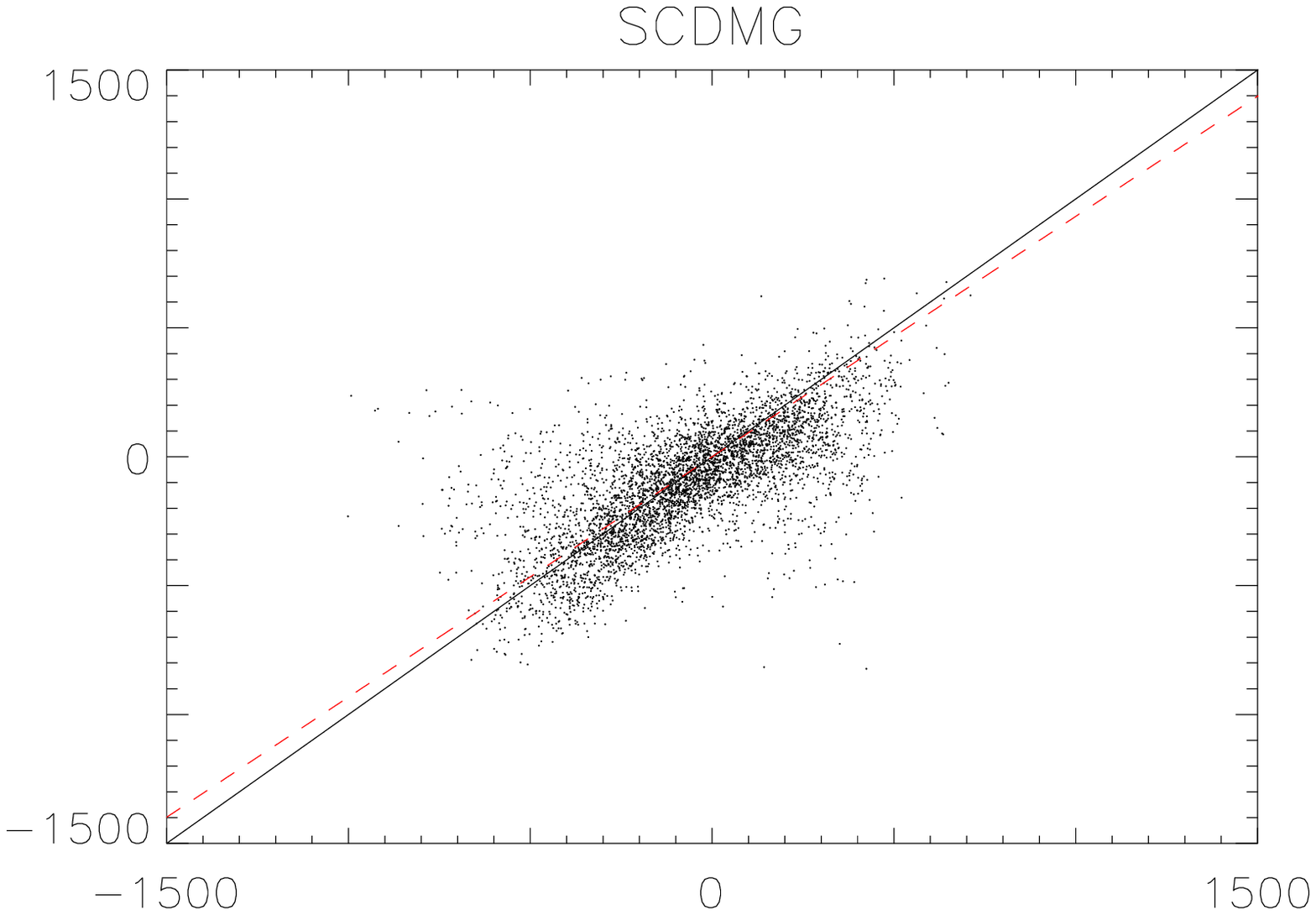}
\caption{Real against predicted velocities for three different cosmologies;
one catalog for each cosmology; plot galaxies out to 60 \hm.}
\label{fig_5}
\end{figure}

Figure \ref{fig_5} shows the real against predicted radial velocities for one
catalog of each cosmology using a constant filter
reconstruction.
% This figure can be directly compared to figure\ref{fig_vvcomp}.
% As in that case, the reconstruction is good in
% general but there is a lot of dispersion in the case of SCDMC. 
% There is no real
% discernible difference between the two reconstruction
% techniques. However, both for figure \ref{fig_vvcomp} and for fig
% \ref{fig_5} we have selected reconstructions that had little
% dispersion. Interestingly, this leads to different catalogs
% being shown in these figures, i.e., the iterative
% method does well on different catalogs than the Fourier Bessel method
% and vice versa. Since both methods are based on linear
% theory, this seems puzzling, but we have also found that the Fourier
% Bessel method is much more sensitive to the degree of smoothing in the
% input redshifts, for example. In general, it seems that a certain size
% of cluster will produce resonance effects in the expansion (see also
% the comments in section \ref{sec_maskcorr}).

Table \ref{tab_errors1} shows the results of applying our line fitting
to the constant filter reconstructions. The parameter $B$ is the slope
of the line averaged over all 10 catalogs in a given cosmology, and
$d$ is the average distance of a point from the line.

\begin{table*}
\begin{center}
\begin{tabular}{lrr} \tableline
\tableline
\multicolumn{3}{c}{LCDM Results} 
\\ \tableline\tableline
Case  & $B \pm \sigma_B$ & $d \pm \sigma_d$
\\ \tableline
$r<120$  &  0.92 $\pm$ 0.08 & 244 $\pm$ 25 
\\ \tableline  
$r<60$  & 0.80 $\pm$ 0.07 & 220 $\pm$ 31 
\\ \tableline  
$60 <r< 120$ & 1.14 $\pm$ 0.22 & 270 $\pm$ 35  
\\ \tableline  
$\delta <1$ & 1.17 $\pm$ 0.15 & 233 $\pm$ 25 
\\ \tableline  
$\delta > 1$ & 0.76 $\pm$ 0.34 & 288 $\pm$ 67 
\\ \tableline  
$-1<\delta<1$, $r<60$   & 0.95 $\pm$ 0.07 & 184 $\pm$ 24 
\\ \tableline \tableline  
 \multicolumn{3}{c}{SCDMG Results} \\ \tableline\tableline
$r<120$  & 0.90 $\pm$ 0.07 & 224 $\pm$ 26
\\ \tableline
$r<60$ & 0.82 $\pm$ 0.07 & 212 $\pm$ 29
\\ \tableline
$60 <r< 120$ & 1.05 $\pm$ 0.19 & 241 $\pm$ 40
\\ \tableline
$\delta <1$ & 0.99 $\pm$ 0.08 & 232 $\pm$ 23
\\ \tableline
$\delta > 1$ & 0.74 $\pm$ 0.18 & 278 $\pm$ 52
\\ \tableline
$-1<\delta<1$, $r<60$ & 0.85 $\pm$ 0.07 & 193 $\pm$ 27
\\ \tableline \tableline
\multicolumn{3}{c}{SCDMC Results} \\ \tableline\tableline
$r<120$  & 1.01 $\pm$ 0.11 & 457 $\pm$ 49
\\ \tableline
$r<60$ & 0.84 $\pm$ 0.12 & 407 $\pm$ 71
\\ \tableline
$60 <r< 120$ & 1.27 $\pm$ 0.13 & 517 $\pm$ 28
\\ \tableline
$\delta <1$ & 1.17 $\pm$ 0.16 & 467 $\pm$ 28
\\ \tableline
$\delta > 1$ & 1.37 $\pm$ 1.51 & 751 $\pm$ 685
\\ \tableline
$-1<\delta<1$, $r<60$  & 0.92 $\pm$ 0.15 & 372 $\pm$ 77
\\ \tableline
\end{tabular}
\end{center}
\caption{Parameters of fitted line and velocity dispersions for three
different cosmologies and 6 different cases. All distances are in
\hm. $B$ is the slope of the fitted line, $\sigma_B$ the dispersion in
the average over all 10 catalogs, d is the average distance of a
point from the fitted line and $\sigma_d$ its dispersion.
\label{tab_errors1}}
\end{table*}

% As before, the slope of the fitted line is very close to unity and the
% dispersion $d$ is
% low (at the same level as in table \ref{tab_errors}). 
% There are two interesting trends when comparing these results to
% those listed in table \ref{tab_errors}:
% The fit is actually closer to unity in most cases, i.e. in
% general, the Fourier Bessel reconstruction will produce the lower
% error. As before, $\sigma_B$ increases quite markedly in the D2 case
% (which also has the worst statistics overall again) but unlike before,
% it also markedly increases in the R2 case. If we exclude the SCDMC D2
% case, the errors will then still be only increased to the level of the
% errors in table \ref{tab_errors}. The SCDMC D2 case indicates that the
% reconstruction essentially fails in this case.

% Therefore, the Fourier Bessel reconstruction is in general better than
% the iterative reconstruction, but it is a bit more sensitive to
% radius. It also is more sensitive to strongly non-linear regions. The
% latter might arise because we did not attempt to correct for
% clusters by collapsing them as was done in chapter 2.

A high-density selection in the Fourier Bessel case is more
damaging to the correlation than a high-radius selection.
Note that in the worst possible cosmology
(SCDMC), the correlation practically ceases in the high density case
but is comparatively normal (by the standards of that universe) for
the high radius case. 
For all cosmologies, there is a clear tendency to overpredict
velocities in high density regions.
% This is similar to the tendency to
% underpredict velocities in the iterative method, only in that case the
% tendency is less dependent on local density.

\begin{figure}
\epsscale{1}
\plotone{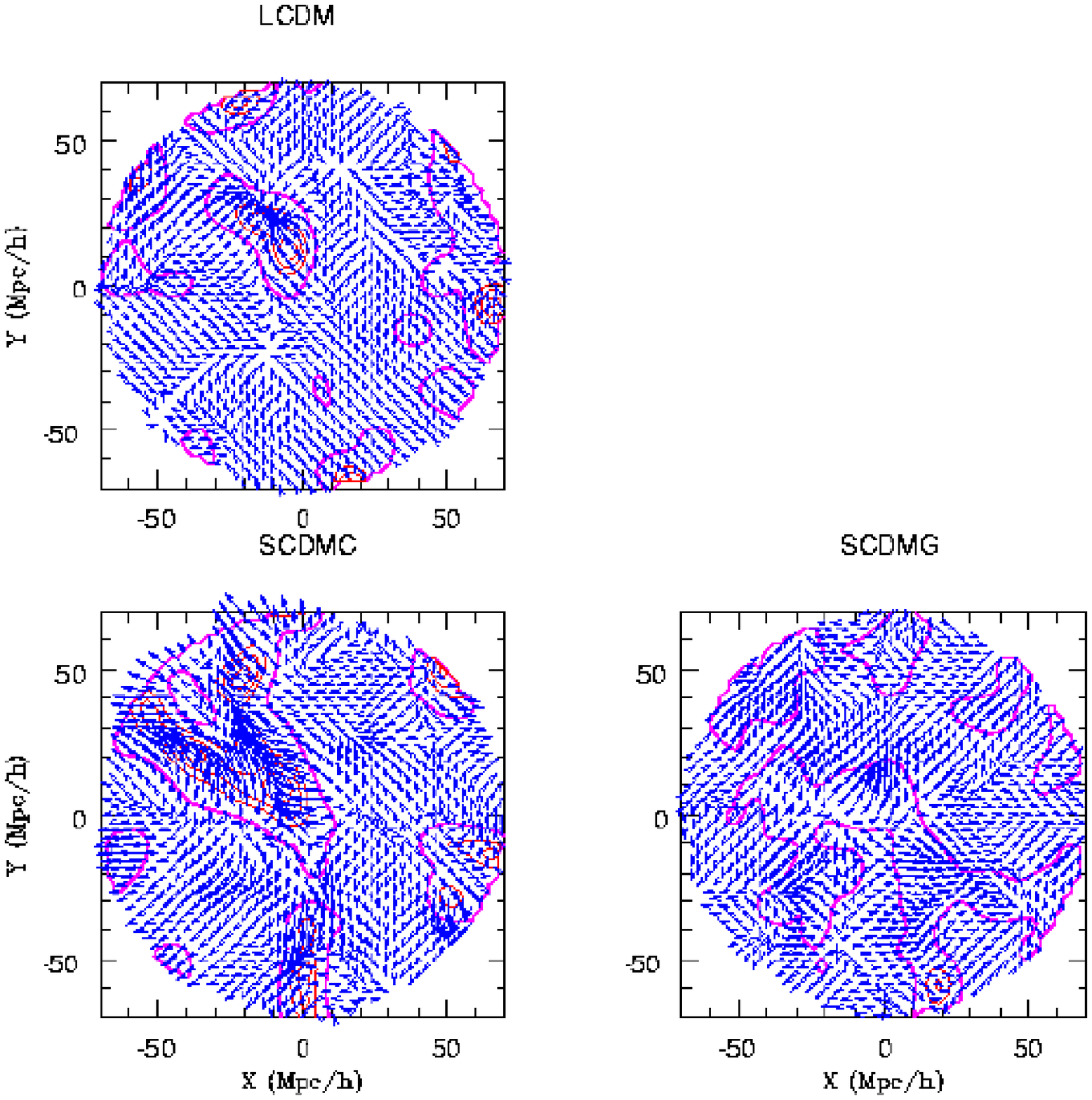}
\caption{$\Delta v = v_{\rm pred} - v_{\rm
real}$ for three different cosmologies (same catalogs as before);
Density levels indicate overdense regions only, bold line is at $\delta
= 0$ and the contours are evenly spaced in 13 steps from $\delta =
1$ to $\delta = 7$. Velocity vectors are drawn on an arbitrary scale.}
\label{fig_6}
\end{figure}

Figure \ref{fig_6} again illustrates the dependence of reconstruction
error on local density:
% (cf also figure \ref{fig_map1}):
for the same
representative reconstructions for each cosmology as in figure \ref{fig_5}, we
plot the reconstructed density field (contours) and the difference
between the reconstructed and the real velocity field (arrows).

The differences are most marked in the SCDMC case and in
all three cases, there is a clear correlation of velocity error with
overdensity. Since the arrows point towards the overdensities, the
infall into clusters is overpredicted.

\begin{figure}
\epsscale{1}
\plotone{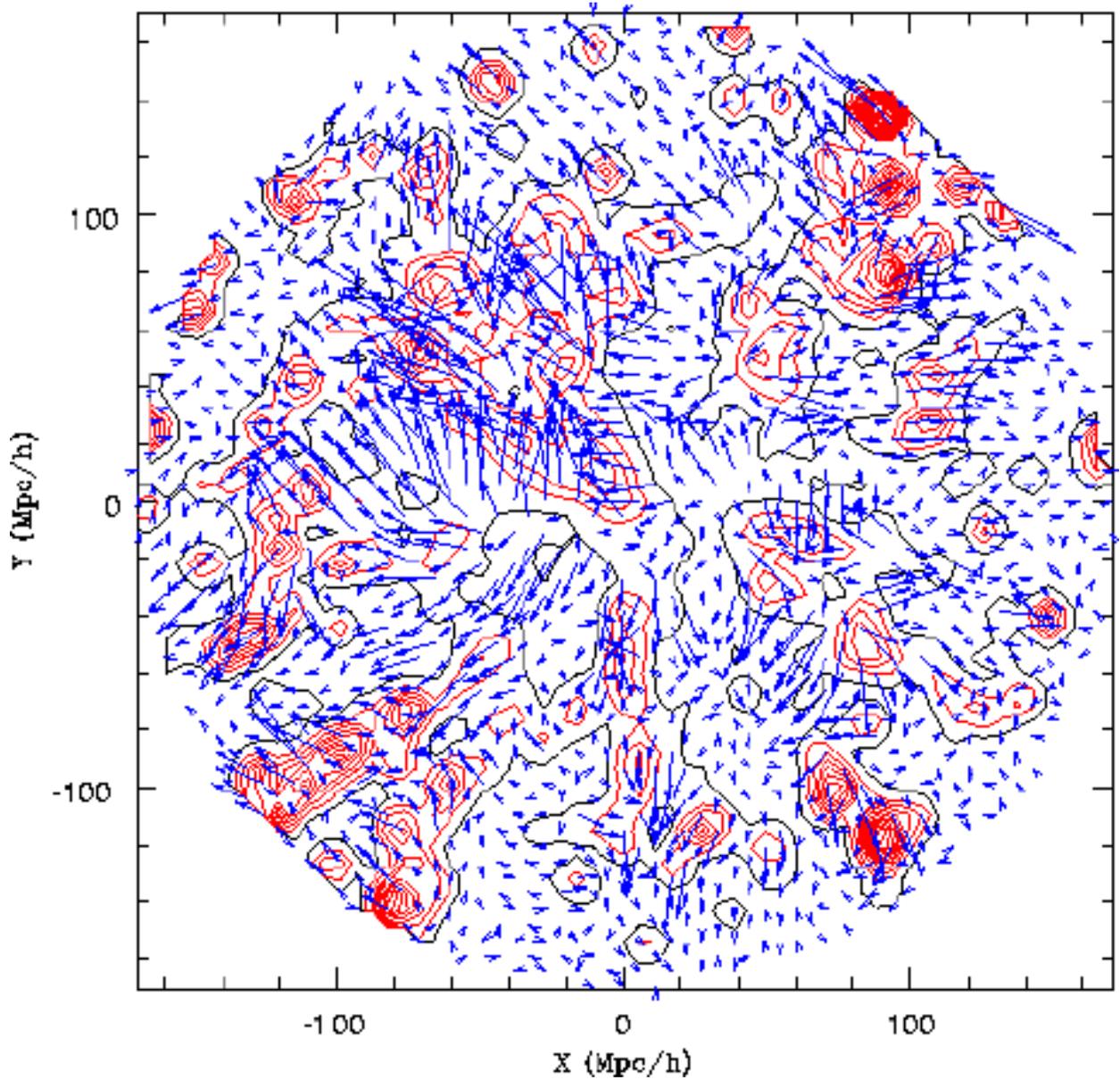}
\caption{$\Delta v = v_{\rm pred} - v_{\rm
real}$ for SCDMC; density levels indicate overdense regions only, bold
line is at $\delta = 0$, thin lines are overdensities from $\delta =
2$ to $\delta = 32$ 
in 16 steps; the smallest velocity difference for this
plane is 18 \kms and the largest 1788 \kms.}
\label{fig_7}
\end{figure}

To test the dependence on radius, we also plot the velocity difference
for the SCDMC case out to 170 \hm (figure \ref{fig_7}). Note that
there is no
indication that the greater shot noise at high radii produces a
systematic effect. Even in those regions, the errors caused by overdense
regions far outweigh any radius dependence. Again, this is a trend
already indicated in table \ref{tab_errors1}.

\section{Local Group Velocity and \bet}
\label{sec_res1}
The velocity field reconstructed by the above method is of particular
interest at the origin: by comparing the reconstructed velocity,
$\vect(v)(0)$, of the Local Group (LG) to the Local Group velocity,
$\vect(v)_{LG}$, measured from the dipole in the CMB we can recover
\bet.

Since only the dipole contributes to the velocity $\vect(v)(0)$ (cf
equation \ref{eq_dip}), we
can set $l_{\rm max} = 1$ for this particular calculation. We also
increase $r_{\rm min}$ to 10 \hm to increase the stability of the
reconstructed LG velocity. If the resolution is too high, the
calculated velocity 
tends to be too easily influenced by small but nearby density
fluctuations.

\begin{figure}
\epsscale{0.8}
\plotone{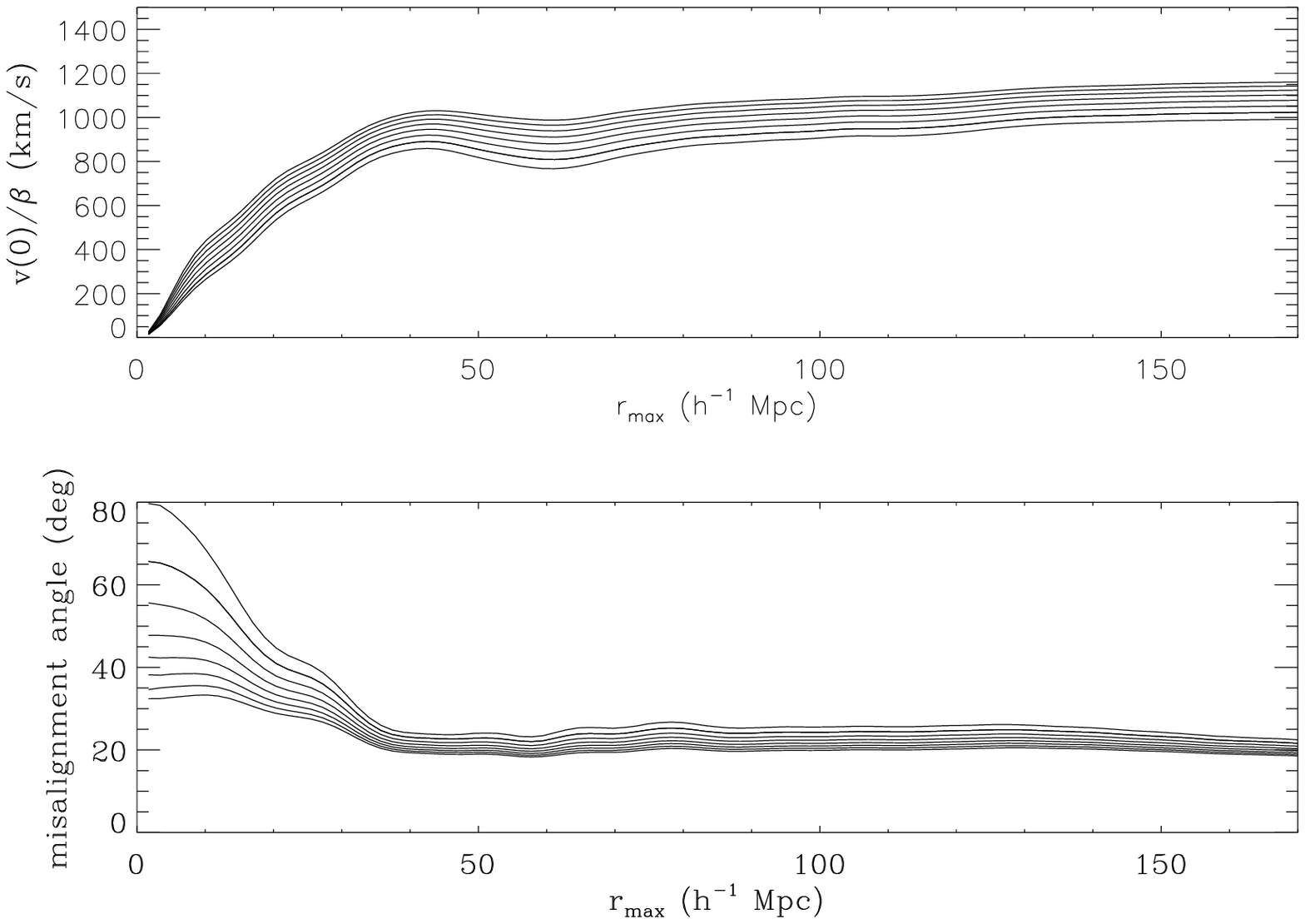}
\caption{Dipole amplitude and misalignment against different
$r_{\rm max}$ for the PSC$z$; note that the velocity is rescaled to
\bet = 1 in the upper panel.}
\label{fig_8}
\end{figure}

Figure \ref{fig_8} shows the measured LG velocity amplitudes for
reconstructions 
using \bet = 0.3 to 1.0 (steps of 0.1) rescaled to \bet = 1.
We plot this against the angle between the reconstructed LG velocity
direction and the CMB 
dipole direction (misalignment angle). 

Note that the reconstructed velocity does
not seem overly sensitive to the \bet assumed in the
reconstruction. The rescaled amplitudes are all in the range between 1000 \kms
and 1100 \kms and hence we can conclude that $\beta \simeq 0.7$. The
very good alignment between the reconstructed velocity and the CMB
dipole indicates that we are satisfactorily sampling the matter that
causes the acceleration. Note also that the convergence of the dipole
amplitude is very good. The Fourier Bessel method works
entirely in the CMB frame and we therefore do not have any signature
from the Kaiser rocket effect (discussed, for example, in \cite{papV}).

Instead of performing a likelihood analysis, in this case we evaluate
systematic effects by calculating the LG velocity for all 30
catalogs to look at the scatter in
\be
(\vect(v)(0) - \vect(v)_{LG} )^2.
\ee
We render this dimensionless and define
\be
\mu^2 = \left( \left(\frac{v}{v_{LG}}\right)^2 - 2
\cos \vartheta \frac{v}{v_{LG}} + 1 \right),
\label{eq_mu}
\ee
where $\vartheta$ is the misalignment angle.  This method may be less
sophisticated than a likelihood analysis (e.g., \cite{me}) but it
makes fewer assumptions about the underlying cosmology. It works on
the simple basis of comparing real data against results from simulated
catalogs.

\begin{figure}
\epsscale{0.8}
\plotone{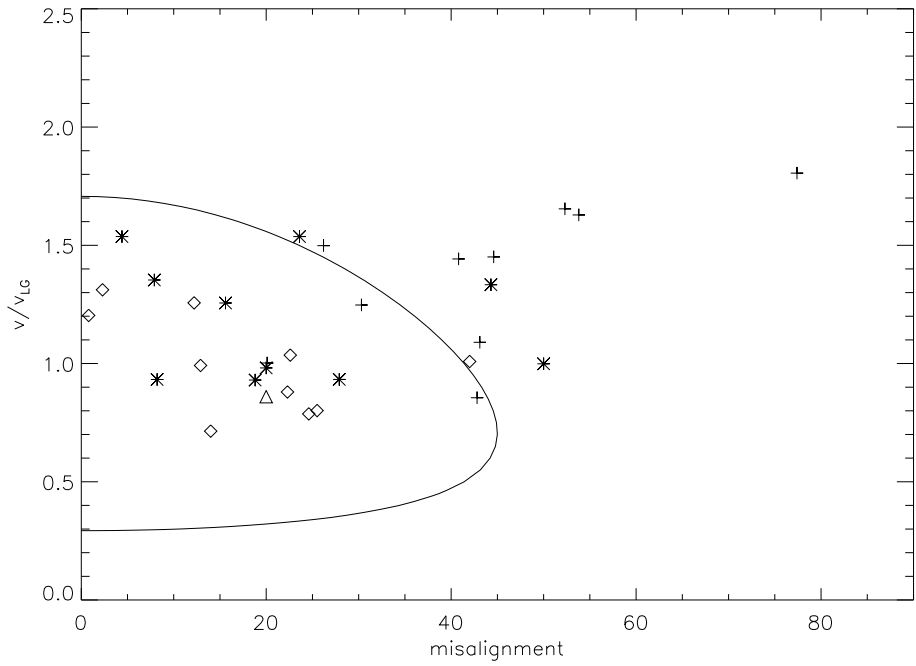}
\caption{Dipole determinations for all catalogs: triangle = PSC$z$,
+ = SCDMC, * = SCDMG, square = LCDM.}
\label{fig_9}
\end{figure}

Figure \ref{fig_9} plots the quantity $v/v_{LG}$ against the
misalignment angle for all simulated catalogs and the PSC$z$
together with contour lines of the surface equation \ref{eq_mu}. The
LG velocity for the PSC$z$ was reconstructed using $\beta = 0.7$ as
was indicated by the above result.  The inner contour encloses 2/3 of
the simulation points and hence it defines our 68 \% confidence limits
on \bet. Therefore, we recover $\beta = 0.7 \pm 0.5$.  This result
seems a lot less restrictive than the one quoted in \cite{me} but it
has to be noted that in that work the results separate the high and
low normalisation cases. In this case, we consider all simulations at
the same time -- drawing together the two different values for \bet in
\cite{me} would give a similarly ill-constrained result.

It is interesting to note that most of the velocity ratios are close
to unity but that the misalignment angle can get quite
large. This is particularly the case for the SCDMC catalogs, which
are (as already mentioned) dominated by large fluctuations at small
radii. This seems to affect the misalignment much more than it does
the amplitude.

\section{Velocity Anisotropy and \bet}
\label{sec_res2}
The density field $\delta^{Re}$ should in general
be perfectly isotropic, i.e., there should be nothing
special about the radial direction.
The radial redshift space distortions in
the density field $\delta^S$, however, depend on \bet so that \bet can
in principle be recovered from analysing those distortions.

We recover the full $\delta^{Re}$ in our method, and if we do so with the
correct \bet, the density field should be isotropic. Therefore, we
simply devise an anisotropy test which will detect radial anisotropies
caused by a reconstruction of the density field using the wrong \bet.
Note that this method tests reconstructed real space velocities and
not the redshift space density field. In order to avoid problems
caused by a small sample (i.e., in order to improve the chance that
the normal anisotropies will average out in the volume considered) we
restrict the analysis to regions where $\delta < 1$.

Increasing \bet in $Z_{lnn'}$ (in equation \ref{eq_short}) 
suppresses structure in the radial direction (since
the redshift space distortions create such structure artificially). If
our assumed \bet is too high, then the recovered $v_r$ will be too
low. We therefore define the parameter
\be
\eta = \frac{1}{3} \frac{ \langle \vect(v)^2 \rangle - \langle
\vect(v) \rangle^2}{\langle v_r^2 \rangle - \langle v_r \rangle^2},
\ee
where the averages are weighted by the local density $(1 + \delta)$.
We expect $\eta$ to pass through unity at the correct \bet.

\begin{figure}
\epsscale{0.8}
\plotone{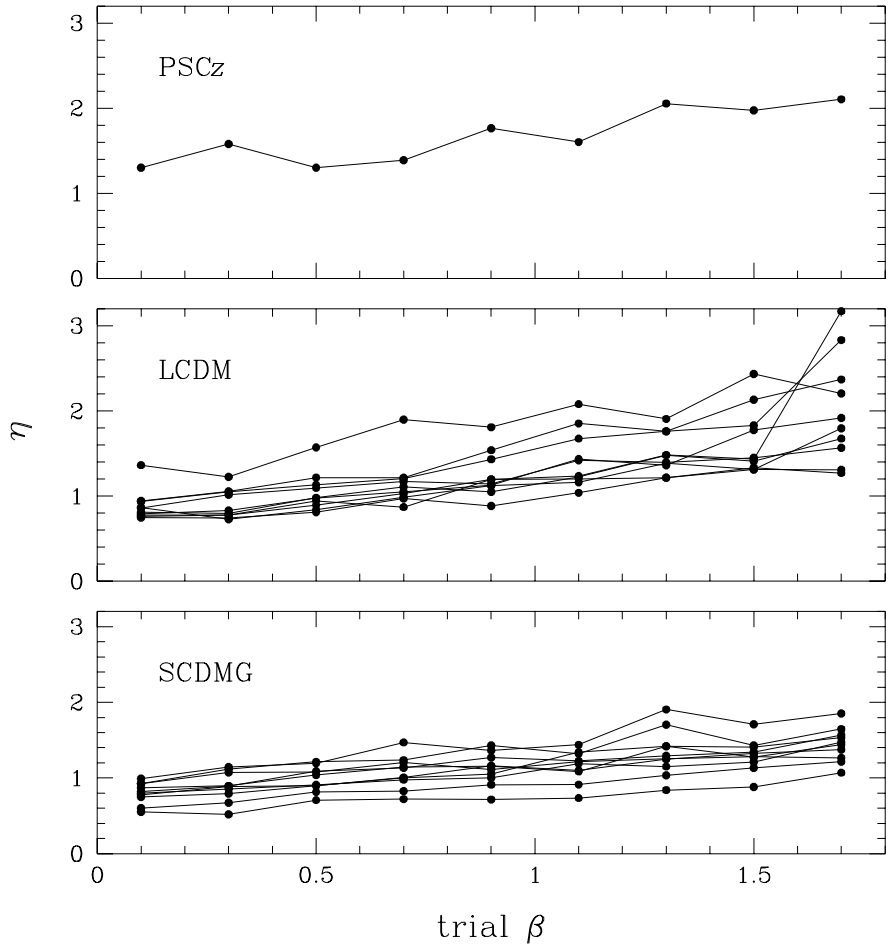}
\caption{Plot of $\eta$ for different trial \bet from PSC$z$, and from
10 SCDM and 10 LCDM catalogs.  The PSC$z$ line lies among the LCDM
values but is almost completely disjoint from the SCDMG values.}
\label{fig_10}
\end{figure}

To calculate the value of $\eta$ for different catalog
reconstructions, we sample the velocity field in a radius of 60 \hm of
the origin with a collection of 800 random points having
$\delta<1$. (We exclude regions with $\delta>1$ since our
reconstructions have larger errors there.) In figure \ref{fig_10}, we
plot $\eta$ against various trial \bet for 10 SCDMG and 10 LCDM
catalogs. We have discounted the SCDMC catalogs for this test
because of their unrealistically large density fluctuations on small
scales (much larger than observed clustering) and partly because of
the high variance in the predicted velocities (see table 1). An
anisotropy test on SCDMC would be too noisy to give useful results.

In general, the scatter is too large for a clear statement about the
\bet at which the lines cross unity. However, we can recognise some
trends. Generally, the SCDMG points (\bet = 1) are lower, i.e., they
cross unity later than the LCDM points (\bet = 0.5) as would be
expected.  This suggests treating $\eta$ as a statistic, and the
values obtained from the 10 SCDMG and 10 LCDM catalogs as samples of
the distribution of $\eta$ in the respective cosmologies.  We find (cf
\ref{fig_10}) that, for all trial \bet, the value of $\eta$ from
PSC$z$ exceeds at least 9 (usually all 10) of the values from PSC$z$.
This is not the case for LCDM. We therefore conclude that SCDM is
excluded at the 90\% confidence level.

It is worth noting that not all \bet =1 models can be rejected on the
basis of this result, since models with more large-scale power will
also have more cosmic variance, increasing the scatter in $\eta$.

\section{Local Density and Peculiar Velocity Fields}
\label{sec_res3}
\begin{figure}
\plotone{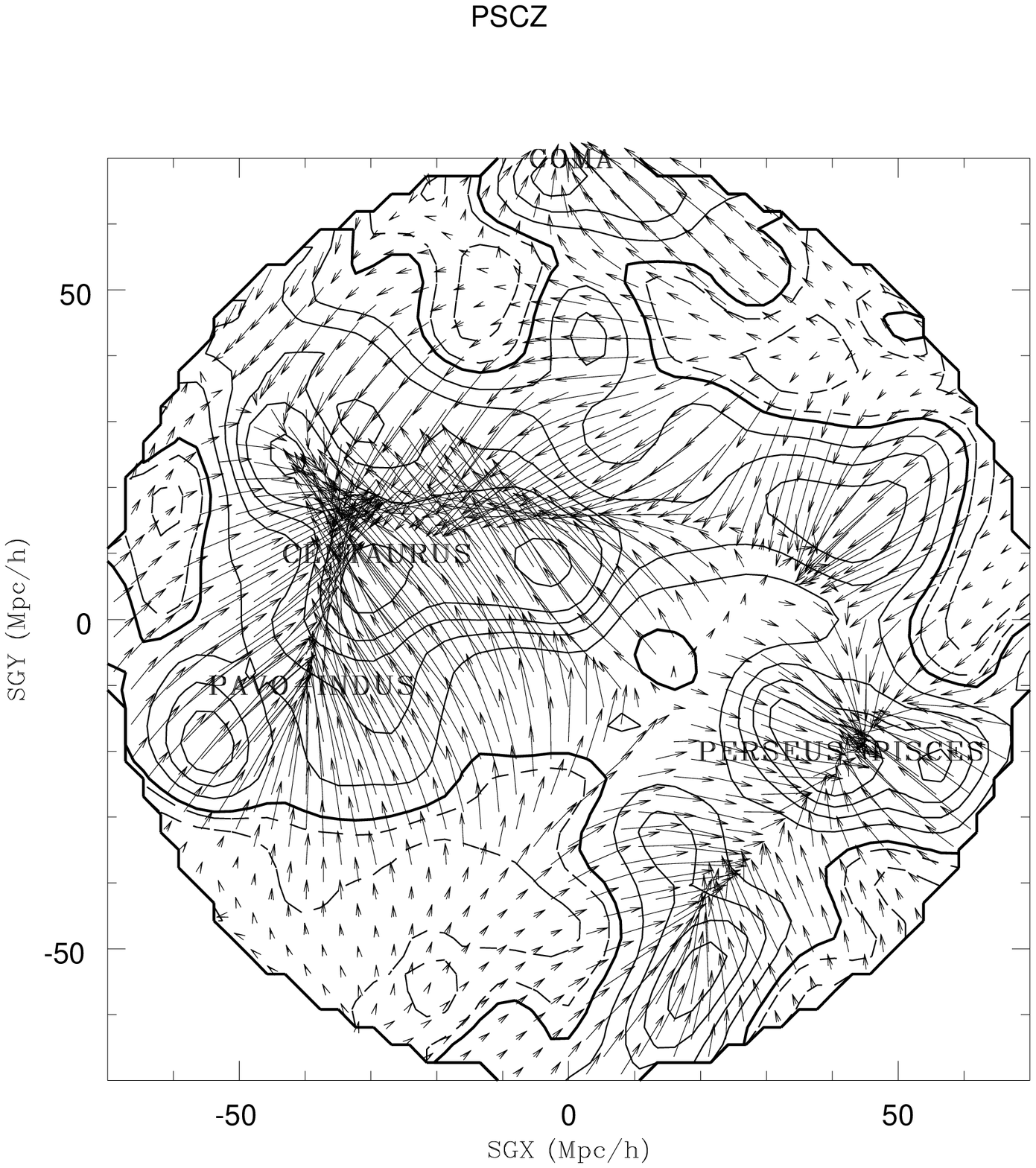}
\caption{ Velocity field of the PSC$z$ survey: radius out to 70
\hm. Contour lines denote density field (fat line = zero density
contrast, broken lines = underdensities, thin solid lines =
overdensities).
%, levels are the same as in figure \ref{velpscz}.
}
\label{fig_11}
\end{figure}

\begin{figure}
\epsscale{1}
\plotone{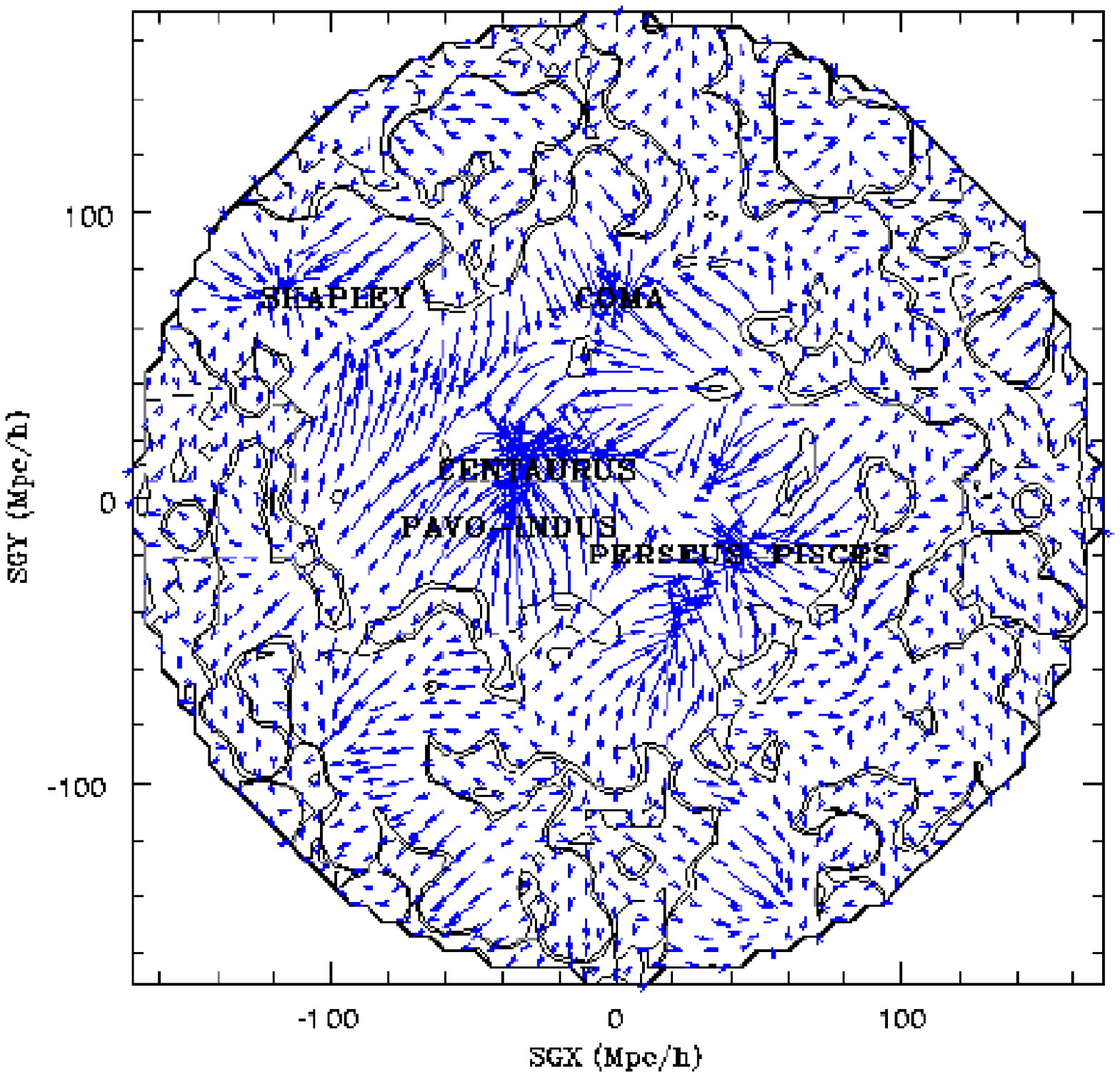}
\caption{Velocity field of the PSC$z$ survey: radius out to 170
\hm.}
\label{fig_12}
\end{figure}

Figures \ref{fig_11} and \ref{fig_12} show the PSC$z$ velocity and
density field in the supergalactic plane out to 70 and 170 \hm
respectively. The general features agree very well with the velocity
field reconstructed in \cite{branch98}. We recover the same continuous
flow from the northern end of the Perseus-Pisces cluster to the GA
region.  However, we also find a back-infall into this region, which
was not observed in \cite{branch98}. This is more apparent in figure
\ref{fig_12}: there is a clear division between the back-infall into
Hydra-Centaurus and the subsequent infall into the Shapley cluster
(-100,90). The highest velocity for any of the back-infalling galaxies
is 494 \kms and therefore above the dispersion level given in table
\ref{tab_errors1}.  However, table \ref{tab_errors1} also indicates
that the infall into high density regions will in general be
overpredicted. Likewise, comparing the real and predicted velocity
maps of the simulated catalogs, we find that the overpredicted
infall makes for a more sharply peaked velocity field in the
reconstructions and hence in about half of the maps for the SCDMG
catalogs, for example, it is easy to identify regions where the
reconstructions show a back-infall into a cluster that does not exist
in the real velocity field. We cannot therefore consider the evidence
conclusive.

\begin{figure}
\epsscale{1}
\plotone{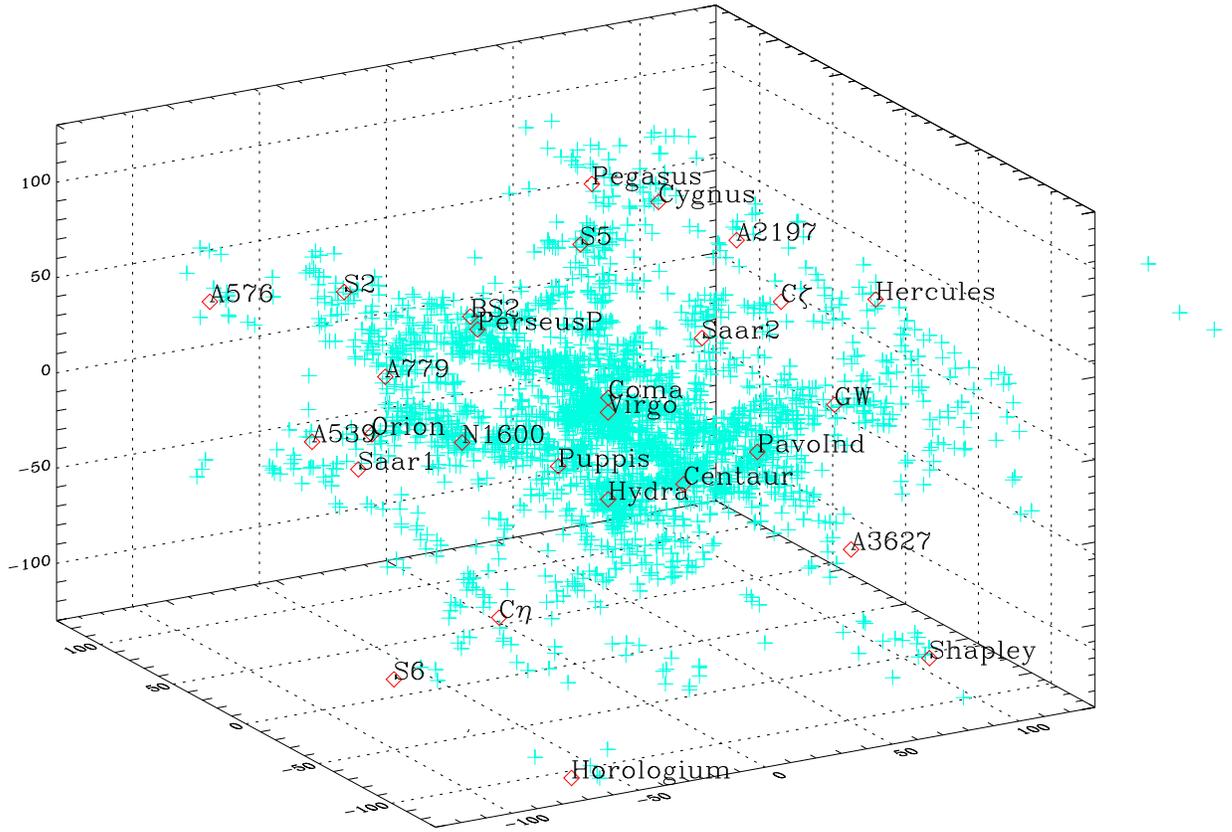}
\caption{Clusters in the PSC$z$ survey: Density level at 1.0.}
\label{fig_13}
\end{figure}

Figure \ref{fig_13} presents the PSC$z$ density field. We have plotted
all those galaxies, for which the local density
$\delta(\vect(r))$ as determined from the density coefficients is
higher than the threshold $\delta = 1.0$ and which are within a
radius of 130 \hm of the observer.

We labelled all the known superclusters as before. The same structures
are identified and it is particularly interesting to note the
extension to Hydra.

\section{Conclusions}

We have corrected the redshift space distortions in the PSC$z$ survey
via the Fourier-Bessel method of FLHLZ with two extensions: a generic
galaxy weight and a matrix correction for the masked zones. The method
was tested extensively on mock catalogs extracted from $N$-body
simulations. It was shown not to have any systematic biases except for
an overprediction of velocities in high density regions. We find 
\begin{itemize}
\item an LG velocity that well reproduces the CMB dipole for \bet =
0.7. We estimate the error on this value from the scatter in the
simulations as $\pm$ 0.5.
\item that an $\Omega = 1.$, cluster normalised and standard CDM
cosmology is ruled out with a confidence of 90 \% from a new
anisotropy test.
\item a flow field in the supergalactic plane that is consistent with
the results of \cite{branch98} apart from an observed back-infall into
the Hydra-Centaurus region. However, we treat this result with caution
since the simulations show spurious features of this type.
\item a density field that shows all the usual clusters and confirms
two of the ones identified by \cite{webster}. There is a possible
conincidence of Saar2, a structure identified in the PSC$z$ by
\cite{veikko}, and the C$\zeta$ cluster of \cite{webster} and likewise
of Saar1 with the Orion and A539 clusters. We also confirm the
existence of A3627 and identify an extension to the Hydra-Centaurus
supercluster.
\end{itemize}

\acknowledgments

I.M.S. and P.S. wish to dedicate this paper to the memory of Veikko
Saar who started this work but was unable to finish it. The authors
are grateful to Bill Ballinger, James Binney, and Radek Stompor for
helpful feedback.  I.M.S. acknowledges financial support from Oriel
College and PPARC, and is particularly grateful to the Sasakawa
Foundation and Tokyo University for funds and hospitality.

\appendix

\section{Dynamical Theory}
\label{app_dyn}
This appendix derives the equations in section \ref{sec_meth} that
express the express the real-space overdensity and velocity fields
under linear theory from survey data via a Fourier Bessel expansion.

We will first introduce the mathematics associated with the distortion
correction. These are independent of the choice of basis functions so
it is only after the derivation that we motivate our particular set of
functions (spherical harmonics and spherical Bessel functions) by the
observation that they are eigenfunctions of the Laplacian operator and
therefore render the velocity calculations particularly easy. We then
present the form of the distortion correction for this basis function
set.

\subsection{Redshift Space Distortions}
\label{sec_t2}

The transformation between redshift space $\vect(s)$ and real space 
$\vect(r)$ is given by
\be
\vect(s) = \vect(r) + \bomega u(\vect(r)),
\label{eq_transform}
\ee
where $u(\vect(r))$ is the radial peculiar velocity at a point
$\vect(r)$ and $\bomega$ is the angular unit vector,
i.e., $\bomega = \hat{\vect(r)}$.

Consider some function $f(\vect(s))$ of the galaxy positions
$\vect(s)$ in redshift space given by a survey.
%What does it mean, though?
Summing over that function at the galaxy positions will be equal to
the integral over all space of the function multiplied with the
the sampling function of the galaxies, i.e.,
\be
\sum_{s_i<R} f(\vect(s)_i) = \int_{4\pi} d\bomega \int_0^R f(s,\bomega) 
\rho^S(s,\bomega) \phi(r) s^2 ds,
\label{eq_first}
\ee
where $\rho^S(s,\bomega)$ is the redshift space density field, $\phi(r)$ is the
selection function (note that it depends on $r$, not $s$, since a galaxy's
inclusion into a flux-limited survey depends on its position in real
space) and $R$ is the maximum radius of the survey in redshift space.
To a zeroth order approximation, this sum will be equal in redshift and
in real space. In the following, we will work out a first order
correction to that approximation by finding the analogous sum in real space.

We now hold $\bomega$ constant since the effect of redshift space
distortions is isolated in the radial component. Mass
conservation requires that
\be
\rho^S(s) s^2 ds = \rho^{Re}(r) r^2 dr,
\ee
where $\rho^{Re}$ is the real space density field,
so we can rewrite the  
right hand side of equation \ref{eq_first}
\be
\int_0^R f(s) \phi(r) \rho^S(s) s^2 ds = \int_0^{R-u} f(s) \phi(r)
\rho^{Re}(r) r^2 dr.
\ee
Note that the boundaries on the
integrals have changed. The maximum radius $R$ is defined in redshift
space and therefore also has to be transformed to real space.
We then expand $f(s)$ in a Taylor
series of the form
\be
f(r+u(r)) = f(r) + uf'(r) + \ldots
\ee
and drop all second order terms, so that
\be
\int_0^R f(s) \rho^S(s) \phi(r) s^2 ds 
= \int_0^{R-u} f(r) \rho^{Re}(r) \phi(r) r^2 dr
 + \int_0^{R-u}
uf'(r) \rho^{Re}(r) \phi(r) r^2 dr.
\label{eq_nosecondorder}
\ee
The first integral on the right hand side has to be split into two
parts
\be
\int_0^{R-u} f \phi \rho^{Re} r^2 dr = \int_0^{R} f \phi
\rho^{Re} r^2 dr + \int_{R}^{R-u} f \phi \rho^{Re} r^2 dr
\label{eq_twoparts}
\ee
since it is a zeroth-order term and the integral between $R$ and $(R-u)$
cannot be neglected (note that we have ceased to give the dependence
on $r$ in all the functions to make the equations neater).
The last term in equation \ref{eq_twoparts} is
first-order, so we can replace $\rho^{Re}$ by 
its average value $\bar{\rho}$, multiply the integrand with $u$,
and evaluate the functions at 
$R$.

The second integral in equation \ref{eq_nosecondorder} is a
first-order term, so we can again replace $\rho^{Re}$ by $\bar{\rho}$ and
integrate by parts. Simplifying, we get
\be
\int_0^R f(s) \rho^S(s) \phi(r) s^2 ds = \int_0^{R} f \phi \rho^{Re}
r^2 dr
- \bar{\rho} \int_0^R f \frac{d}{dr} \left(
u \phi r^2 \right) dr
\ee
Here the first term is the real space expression (analogue of the
redshift space sum in equation \ref{eq_first})
that we were looking for and the second term 
represents the redshift space distortions. 
We can gain some insight into the nature of these distortions by
rewriting this term as
\be
\bar{\rho} \int_0^R f \phi r^2
\left [ u' + \left( \frac{d \ln \phi}{d \ln r} + 2 \right) \frac{u}{r} \right]
dr,
\label{eq_dyn}
\ee
where the terms in the square brackets now describe the various
contributions to the distortions, i.e.,
\begin{itemize}
\item the galaxies' peculiar velocities: the velocity field at a real
space position $\vect(r)$ will be different from the field at
$\vect(s)$ -- hence the $u'$ term.
\item the selection function: there is a change in the selection
function between $r$ and $s$ -- hence the $d\ln\phi/d\ln r$ term.
\item the volume change: a shell at $r$ has a different volume from a
shell at $s$ -- hence the $2u/r$ term.
\end{itemize}
These were first pointed out by \cite{ka87}. Note that of the two
kinds of redshift space distortion discussed in that paper, this
expression only treats the second, linear kind and hence a redshift
space correction on the basis of this expression will only work in the
linear regime.

Now, putting back the $\bomega$-dependence, we have
\begin{eqnarray}
\sum_{s_i<R} f(\vect(s)_i) 
=\int d\bomega \left[ \int_0^R f(r,\bomega) \phi(r) \rho^{Re}(r,\bomega) r^2
dr - \right. \nonumber \\
\left. \bar{\rho} \int_0^R f(r,\bomega) 
\frac{\partial}{\partial r}\left( u(r,\bomega) \phi(r) r^2 \right)
dr. \right]
\label{eq_dynom}
\end{eqnarray}

This represents the general equation for the correction of the redshift
space distortions. We shall find a more specific expression after
deciding on the form of $f$.

\subsection{Peculiar Velocities}
\label{subsec_pecvel}
In linear theory, the velocity is the gradient of the gravitational
potential (see \cite{PE93}, p. 116)
\be
\vect(v) = - \frac{2}{3} \frac{f(\Omega_0)}{\Omega_0 H_0} \nabla \psi.
\ee
Note that $\vect(v)$ is, as usual, a comoving peculiar velocity at the
present time of which $u$ in equation
\ref{eq_transform} is the radial part. 
Poisson's equation states that
\be
\nabla^2 \psi = \frac{3}{2} \Omega_0 \delta_{\rm matter} H_0^2.
\ee
We use the standard approximation $f(\Omega_0) \simeq
\Omega_{0}^{0.6}$, convert the 
density contrast $\delta_{\rm matter}$ 
to the galaxy
density contrast $\delta$, assuming constant bias.
The velocity therefore becomes
\be
\vect(v) = H_0 \beta \vect(\nabla) (\nabla^{-2} \delta),
\label{eq_vvel}
\ee

\subsection{Fourier Bessel Expansion}
\label{sec_dyn}
Since peculiar velocities are our main interest, we want to render the
solution of equation \ref{eq_vvel} as easy as possible. 
It is obvious that the calculation of $\vect(v)$ will be
trivial if we can find an expansion for $\delta$ that is an
eigenfunction of the Laplacian operator. We also require the expansion
to be in spherical coordinates\footnote{Spherical coordinates are given by
$x = r\sin\theta \cos \phi  = r \cos b \cos l$,
$y = r \sin \theta \sin \phi = r \cos b \sin l$, 
$z = r \cos \theta = r \sin b $.
We continue to use $\bomega$ as a shorthand for $\theta,\phi$ - $\phi$
here being the polar angle, not the selection function.}
: firstly because a separation
of the angular and radial parts will concentrate the redshift space
distortions in only one dimension, but also because the decreasing
selection function of a redshift survey will tend to render its volume
spherical.
We therefore choose the Fourier
Bessel expansion
\be
F(r,\bomega) = \sum_{lm} F_{lm}(k) Y_{lm}(\bomega) \int j_l(kr) dk.
\ee
Since the volume of the redshift survey is finite, we approximate the
integral on the right hand side by the sum
\be
F(r,\bomega) =  \sum_{lmn} F_{lmn} Y_{lm}(\bomega) j_l(k_{ln} r),
\label{eq_f}
\ee
i.e., we confine ourselves to $k$-values satisfying certain boundary
conditions on the survey range $R$. There will then be an
orthogonality relation
\be
\int_0^R j_l(k_{ln} r) j_l(k_{ln'} r) r^2 dr =  \delta^K_{nn'} C^{-1}_{ln},
\label{eq_ortho}
\ee
where $\delta^K$ denotes a Kronecker $\delta$. The $Y_{lm}(\bomega)$
are of course orthonormal. 
The values of $C_{ln}$ and $k_{ln}$ depend on the precise boundary
conditions. We follow one common choice, which is to require $\delta =
0 $ at $r > R$ with $\delta$ allowed to be discontinous at $r = R$,
but the logarithmic derivative of the gravitational potential
$d\ln\psi/d\ln r$ required to be continous at that boundary.
In this case, 
\begin{eqnarray}
&&k_{ln} \mbox{ are zeroes of } j_{l-1}(kR)  \\
&&C_{ln}^{-1} = \frac{1}{2} R^3 \left(j_l(k_{ln}R) \right)^2.
\end{eqnarray}
For a discussion of other possible boundary conditions see FLHLZ, appendix A.

If $F(\vect(r))$ was known everywhere, we could simply invert
equation \ref{eq_f} using orthogonality to obtain
\be
F_{lmn} = C_{ln} \int_{V_R} Y^*_{lmn}(\bomega) j_l(k_{ln}r) F(\vect(r)) 
d^3 \vect(r)
\ee
However, $F(r,\theta,\phi)$ is known only with
some position-dependent accuracy $W(\vect(r))$
(such as, for example, the selection function combined with an angular
mask). We then estimate
$F_{lmn}$ in a least-squares sense by minimising
\be
\int_{V_R} \left| F(\vect(r)) - \sum_{l'm'n'} F_{l'm'n'} Y_{l'm'}(\bomega)
j_{l'}(k_{l'n'} r) \right|^2 
W(\vect(r)) d^3 \vect(r).
\ee
Differentiating with respect to the $F^*_{lmn}$ and setting the result
to zero, we obtain
\be
\int_{V_R} F(\vect(r)) Y^*_{lm} j_{l}(k_{ln}r) W(\vect(r)) d^3 \vect(r)
= 
\sum_{l'm'n'} F_{l'm'n'} \int_{V_R} W(\vect(r)) Y_{l'm'} Y^*_{lm}
j_{l'}(k_{l'n'}r) 
j_{l}(k_{ln}r) d^3 \vect(r).
\ee
This is the most general possible expression. We can recover the
$F_{lmn}$ (corrected for the function $W(\vect(r))$) by calculating and
inverting the matrix on the right hand side. However, since the
angular part will have $(2l_{\rm max} +1)^2$ elements and the radial
part $n_{\rm max}(l)$, the matrix will become huge and 
computationally difficult to invert. This problem can be alleviated if
$W$ depends on $r$ only. 
In the following, $W(r) = \phi(r) w(r)$, where $w(r)$ is some weight
function. 
In that case, the integral over
$Y_{lm}Y^*_{l'm'}$ will reduce to Kronecker deltas because of the
orthogonality condition (spherical harmonics are orthonormal on a
sphere) and we obtain
\be
\int_{V_R} F(\vect(r)) Y^*_{lm} j_{l}(k_{ln}r) \phi(r)w(r) d^3 \vect(r)
= 
\sum_{n'} F_{lmn'} \int_{0}^{R} \phi(r)w(r) j_{l}(k_{ln'}r)
j_{l}(k_{ln}r) r^2 dr.
\ee
We can recover the $F_{lmn}$ by multiplying by $P_{lnn'}$ on the left
hand side, where the inverse of $P_{lnn'}$ is defined in
(\ref{eq_refw}).  This will enable us to correct for the selection
function but not for any angular mask.
Mask treatment is discussed in section \ref{sec_maskcorr}.

\subsection{Application of Expansion}

We now use the Fourier Bessel expansion (\ref{eq_exp})
and calculate the version of equation \ref{eq_dynom} valid for this
choice of basis functions. Note that equation \ref{eq_exp} is just a
version of equation \ref{eq_f} with the density contrast field
$\delta$ as the function $F$. 
Equation \ref{eq_exp} together with equation \ref{eq_vvel} leads to
\be
u(r,\bomega) = H_0 \beta \sum_{lmn} \frac{j_l'(k_{ln}r)}{k_{ln}}
Y_{lm}(\bomega) \delta_{lmn},
\label{eq_vel1}
\ee
and
\be
\frac{\partial u}{\partial r} + 
\frac{2u}{r} = H_0 \beta \sum_{lmn} \left( \frac{l(l+1)}{k_{ln}^2
r^2} -1 \right) j_l(k_{ln} r) Y_{lm}(\bomega) \delta_{lmn}.
\ee
For the last step, we have used the Bessel equation to eliminate
$j''(k_{ln}r)$. 

We now choose the function $f(r,\bomega)$ in equation \ref{eq_dynom} as
\be
f(r,\bomega) = w(r) j_l(k_{ln}r) Y^*_{lm}(\bomega),
\ee
where $w(r)$ is a weight function (e.g. $1$ or $1/\phi(r)$ or anything
else as discussed above).
Substituting $\rho = \bar{\rho}( 1 + \delta)$ with $\delta$ given by
equation \ref{eq_exp} into equation \ref{eq_dynom}, we obtain
\begin{eqnarray}
&&\frac{1}{\bar{\rho}} \sum_{s_i<R} w(s_i) j_l(k_{ln} s_i)Y_{lm}(\bomega_i) - 
\sqrt{4\pi} \int
w \phi j_0(k_{ln} r) 
r^2 dr = \nonumber \\
&&\sum_{n'} \delta_{lmn'} 
\int w \phi j_l(k_{ln} r) j_l(k_{ln'} r) r^2 dr \nonumber \\
&&- \beta \sum_{n'} \delta_{lmn'} \int w \phi j_l(k_{ln}r) \left[ 
\left( \frac{l(l+1)}{k_{ln'}^2 r^2} - 1 \right) j_l(k_{ln'} r) 
 + \frac{
j_l'(k_{ln'} r)}{k_{ln'} r} \frac{d\ln \phi}{d \ln r} \right] r^2 dr.
\label{eq_fb}
\end{eqnarray}
Here we have moved the $l=0$ terms to the left hand side.  In a
shorter notation, this becomes equation (\ref{eq_short}).

\subsection{The Local Group Velocity}

The procedure outlined above will allow the reconstruction of the
entire velocity field, but of particular interest to us is the
velocity of the Local Group. However,
our methodology uses polar coordinates which are not defined at
the origin. To solve this problem, we could either simply use points
close to the origin or evaluate the limit by explicitly adding up
the contributions to the acceleration from the surrounding density
field. We have chosen the latter method and our treatment follows
closely that of FLHLZ. 

The LG peculiar velocity is given by
\be
\vect(v)(0) = \frac{H_0 \beta}{4 \pi} \int_{V_R} d^3 \vect(r)'
\delta(\vect(r)') \frac{\vect(r)'}{r'^3}.
\label{eq_dipole}
\ee
Substituting a Fourier Bessel expansion for the density field in the
usual way (cf equation \ref{eq_f}), we obtain 
\be
\vect(v)(0) =\frac{H_0 \beta}{4 \pi} \sum_{lmn} \delta_{lmn}^{Re}
\int_0^{r_{\rm max}}
dr' j_l(k_{ln} r') \int_{4\pi} d\bomega' \bomega'
Y_{lm}(\bomega'),
\ee
where $r_{\rm max}$ is the radius out to which the density field is
considered for the calculation of the LG velocity. We can evaluate the last
integral by noting that
\be
\bomega \cdot \int_{4\pi} d\bomega' \bomega'
Y_{lm}(\bomega') = \frac{4\pi}{3} \delta_{l1}^K \sum_{m=-1,0,1}
Y_{lm}(\bomega),
\ee
(see FLHLZ) 
where $\delta^K$ denotes the Kronecker delta. Hence, only the dipole
term $l=1$ survives. Our choice of boundary conditions
implies $k_{1n}
= \frac{n\pi}{R}$, and we can do the first integral analytically to
obtain
\begin{eqnarray}
\vect(v)(0) &=& \frac{H_0 \beta}{\sqrt{12\pi}} \sum_n 
\left(-\sqrt{2}
\Re(\delta_{11n}) \hat{\vect(x)} + \sqrt{2} \Im(\delta_{11n})
\hat{\vect(y)} + \Re(\delta_{10n})\hat{\vect(z)} \right) \nonumber \\
&&\frac{R}{n\pi} \left( 1 - \frac{R}{n\pi r_{\rm max}} \sin \left(
\frac{ n \pi r_{\rm max}}{R} \right) \right)
\label{eq_dip}
\end{eqnarray}
where $\Re$ and $\Im$ refer to the real and imaginary parts of a
complex number respectively.

\subsection{Computational Comments}
In the code, we calculate the $Y_{lm}$ and $j_{ln}$ in the standard
way using Numerical Recipes (\cite{numrec}) routines. The first
derivative of the $Y_{lm}$ can be calculated using recursion relations
from \cite{AS}, equation 8.5.4.
Since $Y_{l,-m} = Y^*_{lm}$ we need to store the coefficients $F_{lmn}$ only
for $m \ge 0$.

\section{The Wiener Filter}
\label{app_wf}
As shown in Appendix \ref{app_dyn}, the real coefficients
$\delta_{lmn}^{Re}$ and the measured coefficients, say $\zeta_{lmn}$,
are in principle connected in a relationship of the type
\be
\zeta_{lmn} = \sum_{n'} Z_{lnn'} \delta_{lmn'}^{Re},
\ee
where $\zeta_{lmn}$ is a shorthand for the left hand side of equation
\ref{eq_short}. 
However, in the real case, we still have the shot noise to deal with,
i.e., the relationship above has to be amended to
\be
\zeta_{lmn} = \sum_{n'} Z_{lnn'} \left( \delta^{Re}_{lmn'} + \mu_{lmn'} \right),
\label{eq_zeta1}
\ee
where $\mu_{lmn}$ are the Fourier Bessel coefficients of the shot
noise. 

As a first step, we can simplify things by noting that there is no
coupling between different $(l,m)$. Therefore, for a given $(l,m)$, we
can write equation \ref{eq_zeta1} more concisely as
\be
\bzeta = \vect(Z) ( \bdelta + \bmu),
\label{eq_zeta2}
\ee
where $\bzeta, \bdelta, \bmu$ are column vectors of size $n_{\rm
max}(l)$ and $\vect(Z)$ is a $n_{\rm max}(l) \times n_{\rm
max}(l)$ matrix (we have dropped the indices $l$ for clarity).

We now want to derive an operator $\vect(T)$ which will allow us to 
get a good estimate $\bdelta_{\rm est}$ of the real $\bdelta$
\be
\bdelta_{\rm est} = \vect(T) \bzeta.
\ee
As always, we minimise the difference between the
$\bdelta_{\rm est}$ and the
real $\bdelta$, i.e., we minimise
\be
\langle ( \bdelta - \vect(T) \bzeta)^{\dag}
( \bdelta - \vect(T) \bzeta)
\rangle.
\ee
In this and the rest of this section, angular brackets indicate an
average over different realisations of the noise instead of the usual
spatial average.
Differentiating with respect to $\vect(T)$ and setting the result to
zero, we obtain
\be
\vect(T) = \langle \bdelta \vect(\bzeta^{\dag}) \rangle \langle
\bzeta \vect(\bzeta^{\dag}) \rangle^{-1}.
\ee

We know that $\langle \bmu \rangle  = \langle \bmu^{\dag}
\rangle = 0 $ since the noise averages to zero, and so, 
from equation \ref{eq_zeta2} we have
\be
\langle \bdelta \bzeta^{\dag} \rangle = 
\langle \bdelta \bdelta^{\dag} \rangle \vect(Z)^{\dag}
\ee
and
\be
\langle \bzeta \bzeta^{\dag} \rangle = \vect(Z) \left[
\langle \bdelta \bdelta^{\dag} \rangle  + \langle \bmu
\bmu^{\dag} \rangle \right] \vect(Z)^{\dag}.
\ee
We call 
$\langle \bdelta \bdelta^{\dag} \rangle $ the signal
matrix $\vect(S)$ (recognising the normal correlation
function) and $\langle \bmu \bmu^{\dag} \rangle$ 
the noise matrix $\vect(N)$. Hence
\begin{eqnarray}
\vect(T) &=& \vect(S) \vect(Z)^{\dag} \left[ \vect(Z) \left( \vect(S) +
\vect(N) \right) \vect(Z)^{\dag} \right]^{-1} \nonumber \\ 
         &=& \vect(S) \left( \vect(S) + \vect(N) \right)^{-1}
\vect(Z)^{-1},
\end{eqnarray}
which is the form of the Wiener Filter in (\ref{eq_filtform}).

{\vskip 10pt}
The signal and noise matrices are given by 
\be
\vect(S) + \vect(N)
= \langle (\bdelta + \bmu)^{\dag} (\bdelta + \bmu) \rangle
\ee
We return to index notation and therefore obtain
\be
S_{lmnn'} + N_{lmnn'} = \langle \hat{\delta}_{lmn}^{Re}
\hat{\delta}_{lmn'}^{Re} \rangle,
\label{eq_expt}
\ee
where $\hat{\delta}_{lmn}^{Re}$ includes shot noise.
Now, consider our
estimate of the real space density coefficients
\be
\hat{\delta}_{lmn}^{Re} = 
\int_{V_R} \phi(r) w(r) \hat{\delta}^{Re}(r,\bomega) j_l(k_{ln})
Y_{lm}^*(\bomega) P_{ln}.
\label{eq_wfdelt}
\ee
We also know that the expectation value for density fluctuations with
the shot noise taken into account is
given by (cf \cite{bertsch})
\be
\langle \hat{\delta}^{Re}(\vect(r_1)) \hat{\delta}^{Re}(\vect(r_2)) \rangle  = 
\xi(\mid \vect(r_1) - \vect(r_2) \mid) + \frac{1}{\bar{\rho} \phi(r)}
\delta_D^{(3)}(\vect(r_1)-\vect(r_2)),
\label{eq_wffluct}
\ee
where $\delta_D^{(3)}$ describes a three-dimensional Dirac delta
function. The first term in this expression describes the signal, whereas the
second part describes the noise. 
$\xi(\mid \vect(r_1) - \vect(r_2) \mid)$ is the correlation
function of the density field. It depends on distance only, since we
assume isotropy of direction. Note that the cross-term in
equation \ref{eq_wffluct} was dropped since we assume that signal and noise
are uncorrelated.

Using equation \ref{eq_wfdelt} to 
calculate the expectation value equation \ref{eq_expt}
and substituting equation \ref{eq_wffluct}, we can isolate the signal
and noise matrices as
\begin{eqnarray}
S_{lmnn'} &=& \sum_{n_1 n_2} \int_{V_R} d^3 \vect(r_1) d^3 \vect(r_2)
\phi(r_1) \phi(r_2) w(r_1) w(r_2)
j_l(k_{ln_1}r_1) j_l(k_{ln_2} r_2) Y^*_{lm} (\bomega_1) Y_{lm}
(\bomega_2) \nonumber \\
& &\xi(\mid\vect(r_1)- \vect(r_2) \mid) P_{ln_1n} P_{ln_2n'} 
\label{eq_s1}
\\
N_{lmnn'} &=& \sum_{n_1 n_2}\int_{V_R} dr r^2 \phi w^2 
j_l(k_{ln_1}r)j_l(k_{ln_2} r) P_{ln_1n} P_{ln_2n'}.
\end{eqnarray}

The noise matrix can then be worked out explicitly. It does not depend
on $m$.

The signal matrix
can be simplified further by first noting that it cannot depend
on $\phi$ or $w(r)$ since it describes the signal only. We are
therefore free to choose any weight function without loss of
generality and set $w(r) \phi(r) = 1$ in the following.
Additionally, consider the relation between the correlation function
$\xi(\mid
\vect(r_1)- \vect(r_2) \mid)$
and the power spectrum $P(k)$ 
\be
\xi(\mid \vect(r_1)- \vect(r_2) \mid) = \frac{1}{(2 \pi)^3} \int d^3
\vect(k) P(k) e^{-i \vect(k) \cdot (\vect(r_1)- \vect(r_2))}
\ee
and the Rayleigh expansion of the exponential in spherical waves (cf 
\cite{arfken}, p 665)
\be
e^{i \vect(k) \cdot \vect(r)} = 4 \pi \sum_{lm} i^l j_l(kr)
Y^*_{lm}(\bomega) Y_{lm} (\bomega_k).
\ee
This yields
\begin{eqnarray}
S_{lmnn'} &=& \frac{2}{\pi} \int_0^{\infty} dk k^2 P(k)
\int_0^R dr_1 r_1^2 j_l(k_{ln}r_1) j_l(kr_1)
\nonumber \\
& & \int_0^R
dr_2 r_2^2 j_l(k_{ln'}r_2) j_l(kr_2) C_{ln} C_{ln'} 
\end{eqnarray}
since the spherical harmonics are orthogonal. The last integral on the
right hand side can be approximated when the upper limit is set to
$\infty$ (i.e., we assume that the volume inside $R$ is representative
of the universe as a whole), and hence
\be
\int_0^{\infty}
dr_2 r_2^2 j_l(k_{ln'}r_2) j_l(kr_2) C_{ln} C_{ln'}  = \left[ 
\frac{\pi}{2} \frac{1}{k^2} \delta_D^1(k-k_{ln'}) \right].
\ee
We therefore have
\be
S_{lmnn'} \simeq
\frac{2}{\pi} \int_0^{\infty} dk k^2 P(k)
\int_0^R dr_1 r_1^2 j_l(k_{ln}r_1) j_l(kr_1) \left[ 
\frac{\pi}{2} \frac{1}{k^2} \delta_D^1(k-k_{ln'}) \right]
\ee
which can then be reduced to
\be
S_{lmnn'} = C_{ln} P(k_{n'}),
\ee
and again does not depend on $m$.

\end{document}